\documentclass[graybox]{svmult}

\usepackage{mathptmx}       
\usepackage{helvet}         
\usepackage{courier}        
\usepackage{type1cm}        
\usepackage{makeidx}         
\usepackage{graphicx}        
\usepackage{multicol}        
\usepackage[bottom]{footmisc}

\makeindex             

\usepackage{natbib}

\usepackage{amssymb,amsmath}

\newcommand{\ebgiant}{KIC\,8410637}
\newcommand{\sun}{\ensuremath{\odot}}
\newcommand{\msun}{\ensuremath{\,M_\odot}}
\newcommand{\rsun}{\ensuremath{\,R_\odot}}
\newcommand{\rhosun}{\ensuremath{\,\rho_\odot}}

\newcommand{\numax}{\mbox{$\nu_{\rm max}$}}
\newcommand{\Dnu}{\mbox{$\Delta \nu$}}

\newcommand{\muHz}{\mbox{$\mu$Hz}}
\newcommand{\kep}{\mbox{\textit{Kepler}}}

\newcommand{\teff}{\mbox{$T_{\rm eff}$}}
\newcommand{\logg}{\mbox{$\log g$}}

\newcommand{\gdor}{\mbox{$\gamma$\,Dor}}
\newcommand{\dscut}{\mbox{$\delta$\,Scuti}}

\begin{document}

\title*{Asteroseismology of Eclipsing Binary Stars in the \textit{Kepler} Era}
\author{Daniel Huber$^{1,2}$}
\authorrunning{Daniel Huber}
\institute{$^{1}$NASA Ames Research Center, Moffett Field, CA 94035, USA; \email{daniel.huber@nasa.gov} \\
$^{2}$SETI Institute, 189 Bernardo Avenue, Mountain View, CA 94043, USA}
%
%
\maketitle

\abstract*{}

\abstract{Eclipsing binary stars have long served as benchmark systems 
to measure fundamental stellar properties. In the past few 
decades, asteroseismology - the study of stellar 
pulsations - has emerged as a new powerful tool to study the structure and 
evolution of stars across the HR diagram. Pulsating stars in eclipsing binary systems
are particularly valuable since fundamental properties (such as radii and masses) 
can determined using two independent techniques. Furthermore, 
independently measured properties from binary orbits can be used to improve asteroseismic 
modeling for pulsating stars in which mode identifications are not straightforward.
This contribution provides a review of asteroseismic detections in eclipsing binary stars, 
with a focus on space-based missions such as CoRoT and \kep, and empirical tests 
of asteroseismic scaling relations for stochastic (``solar-like'') oscillations.}

\section{Introduction}

Asteroseismology has undergone a revolution in the past few decades. 
Driven by multi-site ground-based observing campaigns and high-precision space-based 
photometry, the number of stars with detected pulsations has increased dramatically, 
and pulsation frequencies and amplitudes are measured with unprecedented 
precision. 
In particular the photometric data provided by the CoRoT and \kep\ space telescopes 
have allowed the application of asteroseismology to stars throughout the HR diagram
\citep[e.g.,][]{gilliland10,michel12,chaplin13b}.

Owing to the relatively large apertures of some
space-based telescopes, however, the majority of stars with high quality asteroseismic 
detections are relatively faint, and hence lack independent observational 
constraints from classical methods such as astrometry or long-baseline interferometry. 
Combining independent observations
with asteroseismology is crucial to advance 
progress in theoretical modeling of observed oscillation frequencies, 
and the validation of asteroseismic relations to derive fundamental stellar properties.
Therefore, the full potential of asteroseismology can only be realized if 
such observations can be combined with model-independent constraints on 
properties such as temperatures, radii and masses.

Eclipsing binaries have long served as benchmark systems for determine fundamental 
properties of stars from first principles. Similar to measuring oscillation 
frequencies, the observation of photometric eclipses and spectroscopic radial velocities 
can be performed for relatively faint systems, as long as spectral lines of the 
components can be successfully disentangled. 
Furthermore, eclipses and pulsations can be measured using the same data.
Thus, asteroseismology of components in eclipsing binary systems promises to be 
a powerful method to improve our understanding of stellar structure and 
evolution.

\section{Principles of Asteroseismology}

This chapter provides a brief introduction into the basic principles of 
asteroseismology. For a more thorough discussion the reader is referred to 
\citet{CD03}, \citet{aerts10}, and \citet{handler13}.

\subsection{Types of Pulsation Modes}

\begin{figure}[t!]
\begin{center}
\resizebox{\hsize}{!}{\includegraphics{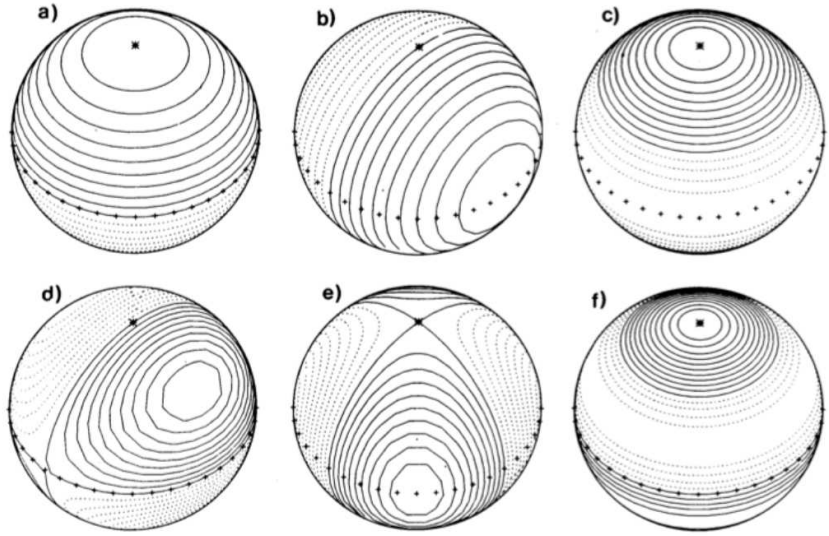}}
\caption{Examples of spherical harmonics used to describe stellar pulsation modes. 
Solid lines show parts of the 
star moving towards the observer, dotted lines parts that move away. The pole of the star is 
indicated by star-signs, the equator by plus-signs. The following cases are shown: 
a) $l=1$, $m=0$; b) $l=1$, $m=1$; c) $l=2$, $m=0$; d) $l=2$, $m=1$; e) $l=2$, $m=2$; f) $l=3$, $m=0$. 
From \citet{CD03}.}
\label{fig:spharm}
\end{center}
\end{figure}

Single pulsation modes in spherically symmetric objects (i.e. in the absence of rapid 
rotation) can be described by 
the quantum numbers $l$, $m$ and $n$. The spherical degree $l$ 
corresponds to the total number of node lines on the surface, and the azimuthal order 
$|m|$ denotes the number of node lines that cross the equator. 
The azimuthal order takes values ranging from $-l$ to $l$ 
(so that $2l + 1$ modes for 
each degree $l$), and is important for rotating stars for which the 
degeneracy imposed by spherical symmetry is broken. The special case of radial pulsations is 
expressed as $l=0$, and corresponds to the star expanding and contracting as a whole 
(sometimes also called the ``breathing mode''). Spherical degrees greater than zero 
are non-radial pulsations, with $l=1$ being 
dipole, $l=2$ quadrupole and $l=3$ octupole modes.
Figure \ref{fig:spharm} shows examples of pulsations modes for several 
configurations of $l$ and $m$. Note that since stars are observed as point sources, 
cancellation effects generally prevent the observation of high degree ($l>3$) modes.
In addition to $l$ and $m$, oscillation modes are further characterized by the radial order 
$n$, the number of nodes along a radius from the surface to the center of the star.

Stellar pulsations can furthermore be separated into two main types: 
pressure modes (p modes) and gravity modes (g modes). 
Pressure modes are acoustic waves propagating through the stellar interior by the 
compression and 
decompression of gas, and the pressure gradient acts as the restoring force. Gravity 
modes correspond to pulsations due to the interplay of 
buoyancy and gravity, and buoyancy acts the 
restoring force. Note that for g modes, no radial ($l=0$) modes exist and the radial 
order $n$ is conventionally counted negative.
The propagation zones of p modes and g modes are generally determined by the position of  
convection zones. Gravity modes are heavily damped in zones where 
convection is unstable, and hence are usually confined to the deep interior for cool stars. 
Pressure modes, on the other hand, propagate in radiative zones, and hence 
are more easily excited to observational amplitudes on the surface. For evolved 
stars, the p-mode and g-mode cavity can overlap, giving rise to so-called 
``mixed modes'' \citep{dziembowski01}. Such modes are of particular importance for studying 
interior properties since they contain contribution from g modes confined to the core, but 
can be observed near the surface. For massive stars with large convective cores 
g modes are more typically observed.

\subsection{Excitation Mechanisms}

Stellar pulsations are excited across a wide range of temperatures and evolutionary 
states in the H-R diagram (see Figure \ref{fig:asterohrd}). The excitation mechanism driving these 
oscillations can be broadly divided into two main types.

\subsubsection{Stochastic Oscillations}

Oscillations in cool, low-mass stars like our Sun are excited by turbulent 
convection in the outer layers of the star \citep[see, e.g.,][]{houdek99}. 
The acoustic energy in the convection zone 
damps and excites oscillation modes stochastically, resulting in mode lifetimes as short as  
a few days in stars similar to our Sun. Finite mode lifetimes cause the peaks in 
the oscillation spectrum to be broadened into a Lorentzian shape, and hence 
such oscillations typically contain no phase information.
Stochastic oscillations are commonly referred to 
as solar-like oscillations, although they are also found in more evolved stars with convective 
envelopes such as red giants. 

\begin{figure}
\begin{center}
\resizebox{\hsize}{!}{\includegraphics{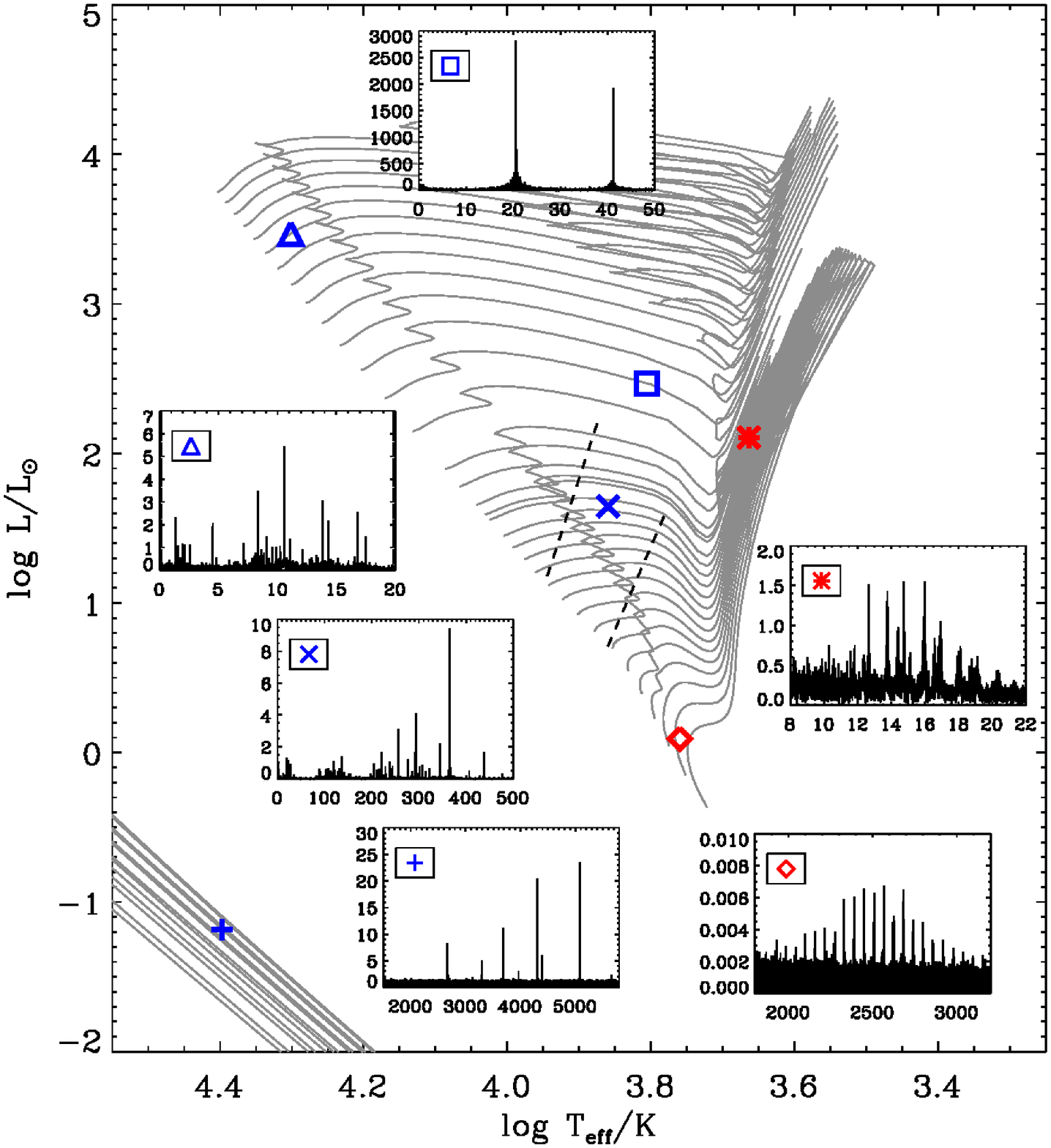}}
\caption{H-R diagram illustrating different types of pulsating stars. Grey lines are 
BaSTI evolutionary tracks \citep{basti} and white-dwarf cooling curves \citep{salaris10}. 
Insets show amplitude spectra in units of ppt versus \muHz\ for 
six typical stars based on 
\textit{Kepler} data: the Sun-like oscillator 16\,Cyg\,B \citep[red diamond,][]{metcalfe12}, 
the red giant 
KIC\,5707854 (red asterisk), the hybrid \gdor-\dscut\ pulsator 
KIC\,11445913 \citep[blue cross,][]{grigahcene10}, 
the hybrid SPB-$\beta$\,Cep pulsator KIC\,3240411 \citep[blue triangle,][]{lehmann11}, 
the RR\,Lyrae star V354\,Lyr \citep[blue square,][]{nemec13} 
and the white dwarf KIC\,8626021 \citep[blue plus symbol,][]{ostensen11}. 
Red and blue symbols denote stars with stochastic and coherent 
oscillations, respectively. 
Dashed lines show the blue and red edge of the \dscut\ instability strip 
\citep{pamyatnykh00}.}
\label{fig:asterohrd}
\end{center}
\end{figure}

Stochastic oscillations are typically of high radial order $n$.
Oscillation frequencies $\nu_{n,l}$ of high radial order $n$ and 
low spherical degree $l$ can be described by the asymptotic theory of stellar 
oscillations \citep{vandakurov68,tassoul80,gough86}, which observationally 
can be approximated as follows:

\begin{equation}
\nu_{n,l} \approx \Delta\nu(n + \frac{1}{2}l + \epsilon) - \delta\nu_{0l} \: .
\label{equ:asymt}
\end{equation}

\noindent
Equation (\ref{equ:asymt}) predicts that oscillation frequencies follow a series of 
characteristic spacings. The large frequency separation
\Dnu\ is the separation of modes of the same spherical degree $l$ 
and consecutive radial order $n$, 
while modes of the different degree $l$ and same radial 
order $n$ are expected to be separated by the small frequency separations $\delta\nu_{0l}$. 
The constant  $\epsilon$ in Equation (\ref{equ:asymt}) is related to the inner and outer turning 
point of the oscillations, and therefore depends on the properties of the surface layers 
of the star. 

In the 
asymptotic theory, the large frequency separation can be shown to be the inverse of twice 
the sound travel time from the surface to the center \citep{ulrich86,CD03}:

\begin{equation}
\Delta\nu = \left(2 \int^{R}_{0} \frac{dr}{c} \right)^{-1} \: ,
\label{equ:dnu_phys}
\end{equation}

\noindent
where the sound speed $c$, assuming adiabacity, is given by

\begin{equation}
c = \sqrt{\Gamma_{1} p / \rho}\: .
\end{equation}

\noindent
Here, $\Gamma_{1}$ is an adiabatic exponent, $p$ is the pressure and $\rho$ is the density. 
For an ideal gas $\rho \propto \mu P / T$, and therefore

\begin{equation}
c \propto \sqrt{T/\mu} \: .
\label{equ:ss}
\end{equation}

\noindent
The sound speed depends on the average internal temperature and chemical composition of 
the gas. For an ideal gas, basic estimates for the central temperature give 
$T\propto \mu M / R$ \citep{KW}, and hence 

\begin{equation}
\Delta\nu \propto \left(\frac{M}{R^3}\right)^{1/2} \: .
\label{equ:delnu}
\end{equation}

\noindent
The large frequency separation is therefore a direct measure of the mean stellar density. 

Figure \ref{fig:asterohrd} shows that the power excess of stochastic oscillations 
has a roughly Gaussian shape, reaching a 
maximum at a certain frequency. This maximum defines the frequency of maximum power (\numax) 
as well as the maximum amplitude of the oscillation, which are related to the driving and 
damping of the modes.
The frequency of maximum power has been suggested to scale with the 
acoustic cut-off frequency \citep{brown91}, which is the maximum frequency below which an acoustic 
mode can be reflected \citep{CD03}:

\begin{equation}
\nu_{\rm ac} = \frac{c}{2 H_{\rm p}} \: .
\end{equation}

\noindent
Here, $H_{\rm p}$ is the pressure scale height which, for an isothermal atmosphere, is given by 
\citep{KW}:

\begin{equation}
H_{\rm p} = \frac{P R^2}{G M \rho} \: .
\end{equation}

\noindent
Using the same approximation as above for an ideal gas and assuming that the temperature can be 
approximated by the effective temperature $T_{\rm eff}$, we have:

\begin{equation}
\nu_{\rm max} \propto \nu_{\rm ac} \propto \frac{M}{R^2 \sqrt{T_{\rm eff}}} \: .
\label{equ:numax}
\end{equation}

\noindent
Actual frequency spectra are considerably more complex than 
described in the above paragraphs. For example, 
frequency separations show variations as a function of radial order which 
depend on the sound speed profile, and hence can be used to infer details on the 
interior structure such as the depth of convection zones or stellar ages 
\citep[e.g.,][]{aerts10}. However, the simple
Equations (\ref{equ:delnu}) and (\ref{equ:numax}) readily relate observables 
to mass and radius, and therefore in principle 
offer a straightforward way to calculate these properties. Importantly, 
these relations are only approximate and require careful calibration 
over a range of evolutionary states. One of the important prospects of 
asteroseismology in eclipsing binary stars is to accurately calibrate these 
scaling relations.

\subsubsection{Coherent Pulsations}

In hotter stars ($\teff\gtrsim 6500\,K$) pulsations can be 
driven by temperature-dependent opacity changes causing 
radiation pressure to continuously expand a star past its equilibrium before contracting 
again under the force of gravity. This heat-engine mechanism (also called $\kappa$ mechanism) 
acting in the hydrogen and helium ionization zones
is effective in a region of the H-R diagram referred to as the classical instability 
strip, which includes pulsators such as $\delta$\,Scuti stars, RR-Lyrae stars, 
and Cepheids (see Figure \ref{fig:asterohrd}). 
Pulsations driven by the $\kappa$ mechanism acting in iron-group elements drive 
pulsations in stars hotter than the classical instability strip, such as 
slowly pulsating B stars and beta Cephei variables. 
Coherent (also called ``classical'') pulsations are phase-stable over 
long timespans of stellar evolution, and typically show significantly 
higher amplitudes than stochastic oscillations (Figure 2).

Coherent pulsations can also be driven by a heat-engine 
mechanism which operates at the base of the outer 
convection zone \citep[``convective blocking'',][]{guzik00}. This mechanism is the 
favored explanation for 
g modes observed in \gdor\ stars, which border stochastically-driven 
oscillators and \dscut\ pulsators in the H-R diagram, as well as H-atmosphere 
white dwarfs (DAV or ZZ Ceti stars). Theoretically the 
$\kappa$ mechanism, convective blocking and stochastic driving should be able to 
excite oscillations simultaneously for stars near the red edge of the instability 
strip. Recent space-based 
observations have established that hybrid \gdor-\dscut\ pulsators are indeed 
common \citep{grigahcene10}, and first evidence for 
hybrid coherent-stochastic pulsators have been found \citep{antoci11}.

\section{The Importance of Eclipsing Binary Stars for Asteroseismology}

\subsection{Asteroseismic Scaling Relations}

Observables of stochastic oscillations
can be trivially related to fundamental stellar 
properties by scaling from the observed values of the Sun. 
For example, Equations (\ref{equ:delnu}) and (\ref{equ:numax}) 
can be rearranged to solve for 
stellar mass and radius:

\begin{equation}
\frac{M}{\mathrm{M}_\odot}=\left(\frac{\nu_\mathrm{max}}{\nu_\mathrm{max,\odot}}\right)^{3}\left(\frac{\Delta\nu}{\Delta\nu_\odot}\right)^{-4}\left(\frac{T_\mathrm{eff}}{\mathrm{T_{eff,\odot}}}\right)^{3/2}\label{eqn7} \: ,
\end{equation}

\begin{equation}
\frac{R}{\mathrm{R}_\odot}=\left(\frac{\nu_\mathrm{max}}{\nu_\mathrm{max,\odot}}\right)\left(\frac{\Delta\nu}{\Delta\nu_\odot}\right)^{-2}\left(\frac{T_\mathrm{eff}}{\mathrm{T_{eff,\odot}}}\right)^{1/2}.\label{eqn8}
\end{equation}

\noindent
The large frequency separation \Dnu\ and the frequency of maximum power \numax\ can be 
easily measured from the power spectrum, providing a straightforward (and in 
principle model-independent) method to determine fundamental properties of stars. 
While matching individual oscillation frequencies to stellar models 
typically yields more precise and detailed information (e.g. on 
initial helium abundances and ages), scaling relations have two important applications. 
First, due to the low 
amplitudes of stochastic oscillations, the signal-to-noise is frequently too low 
to reliably extract a significant number of oscillation modes, and only the average 
large separation can be determined. Second, the measurement of \Dnu\ and \numax\ 
can be performed automatically, and hence 
enables the application of asteroseismology to a large number of stars 
simultaneously. This new era of ``ensemble asteroseismology'' has been recently made 
possible with the launch of space telescopes such as CoRoT and \kep, which provide 
high-precision photometry for thousands of stars.

\begin{figure}
\begin{center}
\resizebox{8.5cm}{!}{\includegraphics{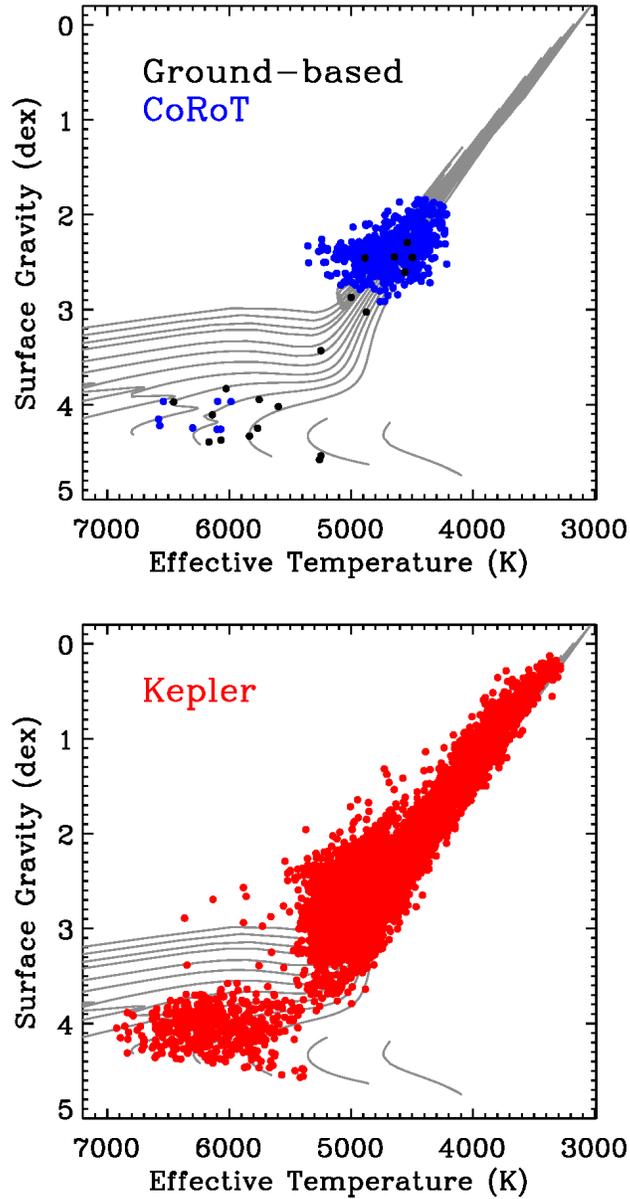}}
\caption{Surface gravity versus effective temperature for stars with detected 
stochastic oscillations as of early 2014. Colors illustrate detections 
by ground-based campaigns (black, top figure), 
CoRoT (blue, top figure) and \textit{Kepler} (red, bottom figure). Grey lines are 
solar-metallicity evolutionary tracks to guide the eye. 
CoRoT red giant detections and \textit{Kepler} detections are taken from the catalogs by 
\citet{hekker09} and \citet{huber14}, respectively.}
\label{fig:history}
\end{center}
\end{figure}

To illustrate this, Figure \ref{fig:history} summarizes the 
detections of stochastic oscillations over the past $\sim$20 years. 
Starting with the first confirmed detection of oscillations in Procyon by \citet{brown91}, 
subsequent ground-based radial-velocity campaigns 
\citep[e.g.,][]{bouchy01,carrier01,kjeldsen05,bedding10} as well as 
observations with early space telescopes such as MOST \citep{matthews07,guenther07} 
yielded detections in 
about a dozen bright stars. Launched in 2006, the CoRoT space telescope delivered 
the first high signal-to-noise detections in main-sequence stars \citep{michel09}, and 
led to the breakthrough discovery of non-radial modes in thousands of 
red giants \citep{deridder09,hekker09,mosser10}. 
The Kepler space telescope, launched in 2009, continued the 
revolution of cool star asteroseismology by filling up the low-mass H-R 
diagram with detections, including dwarfs cooler than the Sun \citep{chaplin11a} and over 
ten thousand red giants \citep{stello13}. The larger number of red giants with 
detected oscillations is due to a combination of two effects: First, oscillation 
amplitudes increase with luminosity \citep{KB95}, making a detection easier 
at a given apparent magnitude. Second, the majority of stars observed by \kep\ are 
observed with 30-minute sampling, setting a limit of
$\logg \lesssim 3.5$ since less evolved oscillate above the 
Nyquist frequency.

It is in particular the large number of oscillating giants which drive the 
need to validate scaling relations. By combining asteroseismic radii and masses 
with temperatures and metallicities, 
stellar ages can be determined for thousands of giant stars, 
opening the door to galactic stellar population studies. Indeed, 
follow-up surveys using multi-object spectrographs such as APOGEE \citep{meszaros13} 
or Stromgren photometry \citep{casagrande14}
have been already dedicated for this purpose. 
The success of this new era of galactic archeology relies on 
our ability to empirically calibrate asteroseismic scaling relations.

Testing the accuracy of 
scaling relations is an active field of research 
\citep[see][for reviews]{belkacem12,miglio13}. 
Theoretical work has shown that both relations typically
hold to a few percent \citep{stello09,belkacem11}, although deviations of the 
\Dnu\ scaling relation by up to 2\% have been reported 
for dwarfs with $M/\msun>1.2$ \citep{white11}. Revised
scaling relations based on model frequencies \citep{white11} and 
extrapolating the measurement of \Dnu\ to higher radial orders \citep{mosser13} 
have been proposed, although some doubt about the applicability of the latter 
revision has been expressed \citep{hekker13b}.
Empirical tests have relied on independently measured properties from 
Hipparcos parallaxes, clusters, and long-baseline interferometry 
\citep[see, e.g.,][]{stello08,bedding11b,brogaard12,miglio11,miglio12b,huber12b,silva12}.
For unevolved stars ($\logg \gtrsim 3.8$) no empirical evidence for 
systematic deviations has yet been determined within the observational uncertainties,  
but for giants a systematic deviation of $\sim 3\%$ in \Dnu\ 
has been noted for He-core burning red giants 
\citep{miglio12b}.

A common limitation is that 
a separate test of the \Dnu\ and \numax\ scaling relation 
relies on an \emph{independent} 
knowledge of stellar mass and radius. Such information is typically 
only available in binary 
systems, either if the masses and radii are measured through an astrometric orbit and 
interferometry, or in doubled-lined spectroscopic eclipsing binaries 
for which absolute masses and radii can be measured. So far, such a test 
has only been possible for three stars: $\alpha$\,Cen\,A, $\alpha$\,Cen\,B, and 
Procyon\,A.

\subsection{Mode Identification and Driving Mechanisms in Intermediate Mass Stars}

Asteroseismology of classical pulsators
has been successfully performed for several decades using
ground-based photometric and spectroscopic campaigns \citep[e.g.,][]{breger00}. 
Intermediate and high mass stars
probe parameter space which are plagued by model uncertainties such as 
convective core-overshooting and the effects of rotation \citep[e.g.,][]{aerts13}. 
Hence, asteroseismology holds 
great promise to improve our understanding of the evolution of such stars.
A serious problem, however, is that classical pulsators often show
complex frequency spectra which do not allow mode identification based 
on simple pattern recognition. In such cases mode identifications rely on 
measuring amplitudes in multicolor photometry or spectroscopic 
line-profile variations, although more recent observations have revealed evidence 
for systematic structure in the frequency spectra of $\delta$\,Scuti stars 
\citep{breger11}\footnote{An important exception are rapidly oscillating Ap stars, 
a class of coherent pulsators showing regularly-spaced high-order 
p modes \citep{kurtz82}.}. 

\begin{figure}
\begin{center}
\resizebox{\hsize}{!}{\includegraphics{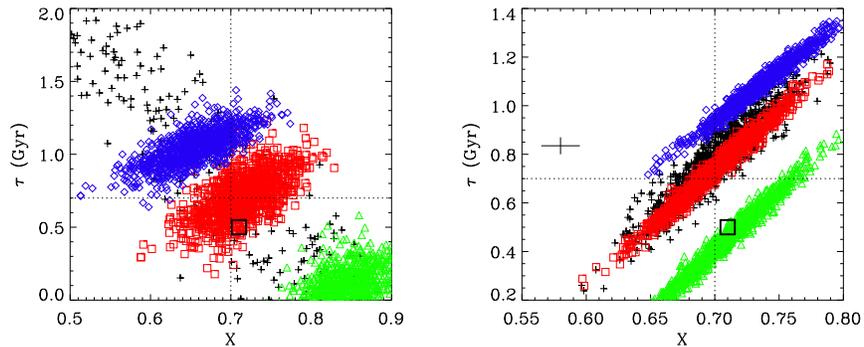}}
\caption{Fitted age versus initial hydrogen abundance for simulated data of a single 
pulsating \dscut\ star (left panel) and an eclipsing binary including a 
\dscut\ star (right panel). Symbols denote simulations including no pulsation mode 
(black plus signs), including a pulsation frequency with a correct mode identification 
(red squares), including a pulsation frequency with incorrect spherical degree 
(blue diamonds), and a a pulsation frequency with incorrect radial order (green 
triangles). Dashed lines mark the exact 
input solution, the box show the initial guess values, and the error bar mark the 
input uncertainties for the eclipsing binary solution. From \citet{creevey11}.}
\label{fig:creevey}
\end{center}
\end{figure}

Classical pulsators in eclipsing binary stars offer the possibility to alleviate this 
problem.
\citet{creevey11} found that eclipsing binaries
constrain fundamental properties 
of a star equally well or better than what is possible for a pulsating 
\dscut\ star with correct mode identification (Figure \ref{fig:creevey}). 
Conversely, the correct mode identification can be 
inferred by comparing solutions assuming different mode identifications, and 
identifying those solutions which yield the best match to the binary constraints. 
Once the mode identification is secured, interior properties 
(such as convective core overshoot or mixing length parameter) and be further constrained 
using the pulsation frequencies.

Eclipsing binary systems may also contribute to addressing
the long-standing question of driving mechanisms near 
the red edge of the instability strip. 
Stochastic oscillations in \dscut\ and 
\gdor\ stars have been predicted theoretically, yet little 
conclusive observational evidence for hybrid oscillators has yet been found. 
Pulsating stars with precisely determined fundamental properties may be
key to understand the interplay between convection and driving of pulsations near 
the red edge of the classical instability strip.

\subsection{Tidally Induced Pulsations and Eccentric Binary Systems}

Multiple star systems offer the possibility to study gravity modes
driven by tidal interactions \citep[dynamical tides,][]{zahn75}, which 
are particularly prominent in eccentric binary systems \citep[e.g.,][]{lai97}. 
Tidal interactions
can also be used to infer properties such as eccentricities, masses and inclinations 
of binary stars and are of importance for planet formation, 
for example for migration theories of hot Jupiters
through high eccentricity migration and tidal circularization \citep[e.g.,][]{winn10}.

Observations of dynamical tides
have long been hampered by the required high precision
photometry and continuous coverage to observe these effects over many 
orbital periods. \kep\ changed this picture by observationally confirming a 
new of class of eccentric binaries with tidally 
induced brightness variations at periastron passage, also known as ``heartbeat''
stars \citep{thompson12}. 
The prototype, KOI-54, consists of two nearly equal mass A stars in a highly 
eccentric orbit \citep{welsh11} with tidally induced pulsations that may be locked into 
resonance with the binary orbit \citep{fuller12,burkart12}.
The discovery of KOI-54 started a new era of 
observational ``tidal asteroseismology''.

While tidal interactions carry a large amount of information about binary systems,
challenges remain which can be addressed if constraints from 
eclipses are available. 
For example, the precise frequencies, amplitudes, and phases of tidally excited 
oscillations may provide a 
wealth of information on tidal damping mechanisms. Independently measured radii and 
masses of the components provide a possibility to accurately model the stellar components 
and calibrate theories of tidal dissipation.

\section{Giant Stars}

Eclipsing binary systems provide several powerful possibilities to 
study the structure and evolution of giant stars, for example through 
the observation of chromospheric eclipses. Due to the large 
oscillation amplitudes and the presence of mixed modes, red giant stars also have large 
potential for asteroseismic studies.

\subsection{Oscillating Giants in Eclipsing Binary Systems}

\begin{figure}
\begin{center}
\resizebox{\hsize}{!}{\includegraphics{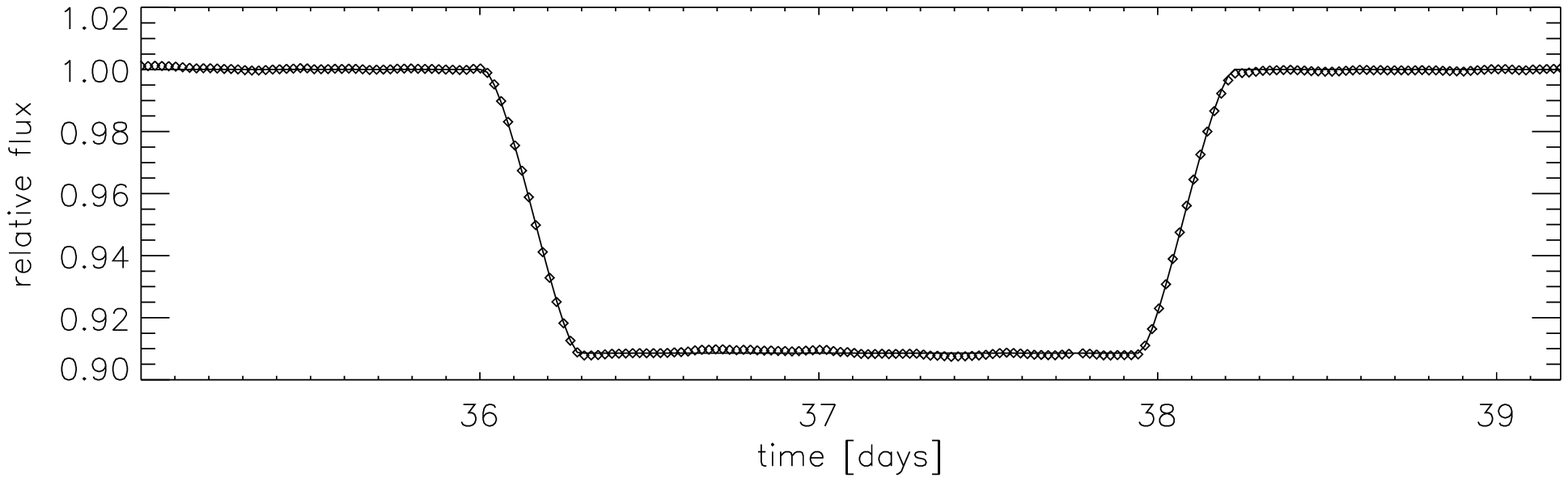}}
\resizebox{\hsize}{!}{\includegraphics{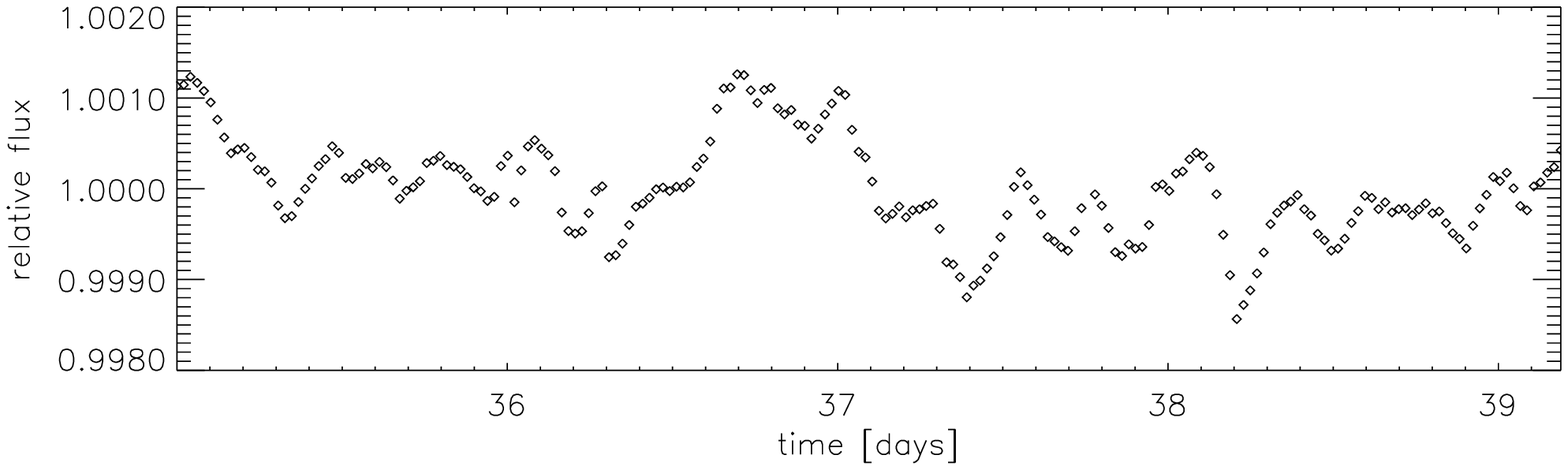}}
\caption{Top panel: \textit{Kepler} discovery light curve of KIC\,8410637, 
the first eclipsing binary with an oscillating red giant component. 
Bottom panel: Light curve after correcting an eclipse model, revealing the 
oscillations in the red giant primary. From \citet{hekker10}.}
\label{fig:giantlc}
\end{center}
\end{figure}

The first detection of an oscillating giant in an eclipsing binary system was 
presented by \citet{hekker10} using \kep\ data. Figure \ref{fig:giantlc} shows 
the discovery light curve of \ebgiant\ (TYC 3130-2385-1, $V=11.3$) based on the first 
30\, days of \kep\ data, revealing a total eclipse (top panel) and 
stochastic oscillations (bottom panel). 
The presence of 
oscillations during the eclipse pointed to a smaller object being occulted by a 
red giant. Since only a single eclipse was observed, the orbital period of the system was 
initially unknown, but constraints through the luminosity and radius ratio 
suggested a F-type main-sequence star as the secondary component. 

\begin{figure}
\begin{center}
\resizebox{9cm}{!}{\includegraphics{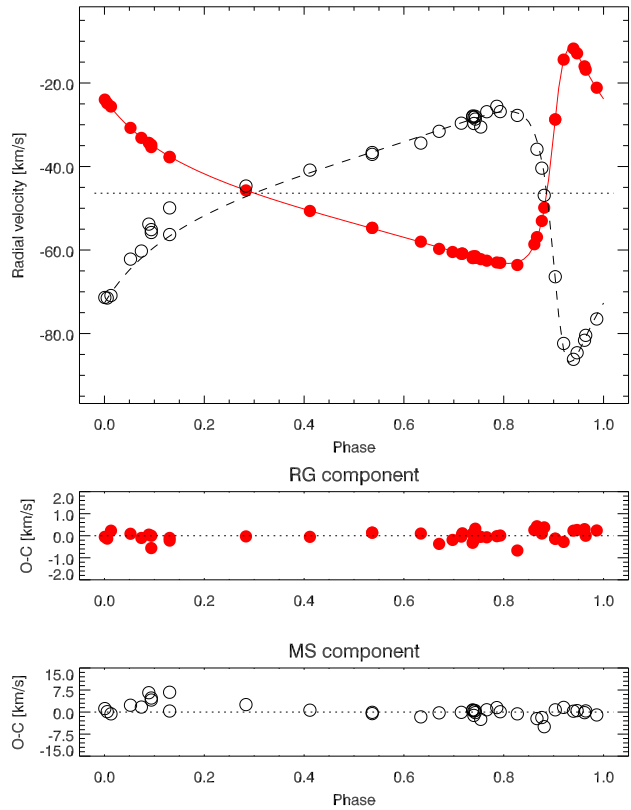}}
\caption{Top panel: Radial velocities phased with the 408 day orbital period of the 
red giant primary (red filled circles) and main-sequence secondary (black open circles) 
component of \ebgiant. 
Bottom panels: Residuals of the best-fitting model solution. \ebgiant\ is the first 
eclipsing binary with a component showing stochastic oscillations and 
a measured double-lined spectroscopic orbit. From \citet{frandsen13}.}
\label{fig:giantrvs}
\end{center}
\end{figure}

\citet{frandsen13} presented an complete orbital solution of \ebgiant\
based on nearly 1000 days of \kep\ data, an extensive 
radial-velocity campaign and multi-color 
ground-based photometry.
Figure \ref{fig:giantrvs} shows the radial-velocity solution, which combined with the 
extended \kep\ dataset spanning three primary and three secondary eclipses was used 
to derive a dynamical solution of the system, with an orbital period of 408 days 
and an eccentricity of 0.6.
The radius and mass of the components were measured to be 
$R_{\rm RG}=10.74\pm0.11\rsun$ and 
$M_{\rm RG}=1.56\pm0.03\rsun$ for the red-giant primary, as well as 
$R_{\rm MS}=10.74\pm0.11\rsun$ and 
$M_{\rm MS}=1.56\pm0.03\rsun$ for main-sequence secondary. 
The exquisite precision of the absolute radii and masses make 
\ebgiant\ an extremely interesting object to test asteroseismic scaling relations for 
evolved stars.

Table 1 compares the solution by \citet{frandsen13} to the values from 
scaling relations presented by \citet{hekker10}. While 
the values are in reasonable agreement, 
the uncertainties on the seismic mass and radius are fairly large since 
the analysis by \citet{hekker10} was based on only 30 days of 
\kep\ data.
To test whether this difference is significant, the analysis was repeated using 
data from Q0-16 (1350 days) with the method by \citet{huber09}. 
The resulting seismic values are $\numax=46.2\pm1.1\muHz$ and $\Dnu=4.634\pm0.012\muHz$, 
in good agreement with the values by \citet{hekker10} but with significantly 
reduced uncertainties (factor of 2 in mass and radius). 
Combining these values with the effective temperature by \citet{frandsen13} 
yields excellent agreement (within 0.003\,dex) in \logg\ (see Table 1). 
However, the density is 
underestimated by $\sim$7\% (1.8\,$\sigma$, taking the uncertainty in the 
seismic and dynamical density into account), which results in an 
overestimate of the radius by $\sim$9\% (2.7\,$\sigma$) and mass by $\sim$17\% (1.9\,$\sigma$).
Note that a $\sim$7\% difference in density is a factor 7 larger than the typical
formal uncertainty on the seismic density from the measurement 
of $\Dnu$ using different methods \citep{hekker12}.
Table 1 also lists radius and mass estimates 
using the recently proposed corrections to the $\Dnu$ scaling relation 
by \citet{white11} and \citet{mosser10}. Both corrections 
reduce the differences to $\sim$4\% in density, 
$\sim$5\% in radius, and $\sim$10\% in mass. 

An important piece of information for \ebgiant\ is whether the primary
is a He-core burning red clump star or still ascending the red-giant 
branch. While an asteroseismic determination of the evolutionary state based on 
gravity-mode period spacings
\citep{bedding11,beck11} is still pending, \citet{frandsen13} argued 
that a red-clump phase of the 
primary is more likely based on the 
comparison of the derived temperature with isochrones. 
However, relative corrections to the 
\Dnu\ scaling relation between red-clump and RGB stars tend to 
increase the seismic mass and radius \citep{miglio12b}, hence resulting in 
even larger differences with the orbital solution.
Additionally, as pointed out by \citet{frandsen13}, the small periastron 
distance would imply that the system may have undergone significant mass transfer 
when the primary reached the tip of the RGB. 

\begin{table*}
\begin{small}
\begin{center}
\caption{Fundamental properties of the red giant component in the eclipsing binary 
system 
KIC\,8410637 from an orbital solution and from asteroseismic scaling relations.}
\begin{tabular}{l | c | c c c c}        
\hline 
Parameter    & RVs+EB & \multicolumn{4}{c}{Asteroseismic Scaling Relations}	\\       
 			 & Frandsen et al. & Hekker et al. & Q1--16$^{*}$ & Q1--16+White$^{\ddag}$ & Q1--16+Mosser$^{\ddag}$	\\
\hline		
\teff\ (K)	 &						$4800\pm80$				& $4650\pm80$ 		& $4800\pm80$ 		& $4800\pm80$ 		& $4800\pm80$  \\		
$R$ (\rsun)&						$10.74\pm0.11$			& $11.8\pm0.6$ 		& $11.58\pm0.30$ 	& $11.23\pm0.29$	& $11.31\pm0.29$ \\		
$M$ (\msun)&						$1.56\pm0.03$			& $1.7\pm0.3$		& $1.83\pm0.14$		& $1.72\pm0.13$	& $1.74\pm0.13$  \\		
\logg\ (cgs)&						$2.569\pm0.009$			& ---				& $2.572\pm0.011$	& $2.572\pm0.011$	& $2.572\pm0.011$  \\
$\rho$ ($\rhosun \times 10^{3}$)&	$1.259\pm0.046^{\dag}$	& ---				& $1.1765\pm0.0061$	& $1.2132\pm0.0063$	& $1.2047\pm0.0061^{\dag}$ \\
\hline
\end{tabular} 
\label{tab1} 
\end{center}
\flushleft 
$^{*}$ For asteroseismic solutions based on Q1-Q16, solar reference values of  
$\nu_{\rm max,\sun}=3090\pm30\,\muHz$ and $\Delta\nu_{\sun}=135.1\pm0.1\,\muHz$ were 
used \citep{huber11b}. \\
$^{\ddag}$ Based on scaling relations corrections proposed by \citet{white11} and 
\citet{mosser13}. \\
$^{\dag}$ Calculated from mass and radius.
\end{small}
\end{table*}

Clearly, larger samples are required to 
determine whether the differences found for \ebgiant\ may be systematic. 
Fortunately, the \kep\ offers a 
goldmine of over 15,000 oscillating giant stars which have been used 
to identify additional systems.
\citet{gaulme13} crossmatched the \kep\ 
eclipsing binary catalog \citep{prsa11,slawson11,matijevic12}
with red giants classified in the Kepler Input Catalog \citep{brown11} to 
identify 12 new candidate eclipsing binary systems with oscillating red giants. 
The red giant components 
in the sample span a large range in evolution, making this sample promising to extend tests of 
asteroseismic scaling relations. Radial velocity follow-up to confirm these systems 
as genuine eclipsing binaries with oscillating giants and to measure absolute 
radii and masses are currently underway.

\subsection{Oscillating Giants in Eccentric Binary Systems}

\begin{figure}
\begin{center}
\resizebox{\hsize}{!}{\includegraphics{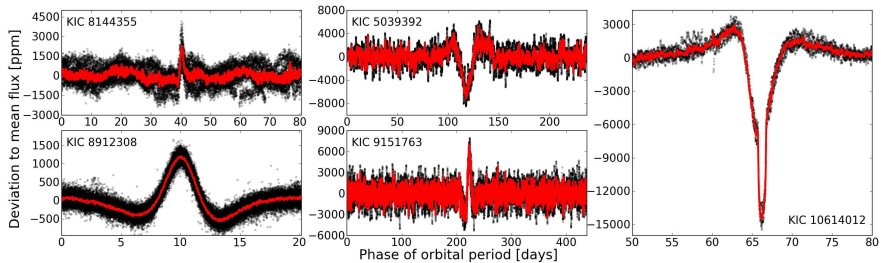}}
\caption{\textit{Kepler} phase curves of five eccentric binary systems with oscillating 
red giants displaying 
gravitationally induced brightness changes near periastron passage (``heartbeat'' stars). 
The shape and amplitude of the distortion depends on the orbital properties 
(e.g. eccentricity, argument of periastron, inclination) and masses of the system.
From \citet{beck13}.}
\label{fig:heartbeats}
\end{center}
\end{figure}

The first oscillating giants in heartbeat systems were presented by \citet{beck13}, 
who confirmed the discovery of 18 Kepler systems through radial-velocity follow-up.
Figure \ref{fig:heartbeats} shows five examples of 
phased heartbeat light curves. The shape, length and amplitude of the 
light distortion depends on the orientation and inclination 
of the orbit, as well as the masses of the components. 

\citet{beck13} presented a detailed study of
KIC5006817, a system with an orbital period of 95 days, eccentricity of 0.7, and 
an orbital inclination of 62 degrees. While the secondary is too faint to be 
detected in the spectra, the primary is relatively unevolved RGB star  
($R=5.84\rsun$, $M=1.49\msun$) showing a high signal-to-noise 
asteroseismic detection. 
The asteroseismic analysis allowed a measurement 
of the stellar inclination axis from rotationally split dipole modes \citep{gizon03} of 
$77\pm9$ degrees. 
The larger value compared to the orbital inclination and the comparison of the 
rotation period inferred from rotational splittings to the orbital period were 
interpreted as evidence that the system has not yet tidally synchronized.

The analysis of KIC\,5006817 yielded several surprising results. First, 
the data show an apparent absence of a signal due to 
Doppler beaming, a periodic increase and decrease in intensity mostly due to the 
radial velocity shift of the stellar spectrum relative to the photometric bandpass. 
Second, the gravity darkening derived from the light curve model disagrees with 
empirical and semi-empirical gravity darkening values. 
While the former may be related to the difficulty of detrending data when the 
orbital period is similar to the length of a Kepler observing quarter, \citet{beck13} 
conclude that the latter 
likely implies a revision of commonly accepted gravity darkening exponents for 
giants (assuming that the derived properties of the red giant are correct).

While the absence of eclipses and the spectroscopic non-detection of the secondary 
precluded an independent measurement of radii and masses of both components, 
heartbeat systems such as KIC\,5006817 allow insights into the dynamical evolution of 
eccentric binary systems with evolved stars. For example, \citet{beck13} find tentative 
evidence that systems with higher larger red-giant primary radii have
longer orbital periods, indicating that some of these systems may form the 
progenitors of cataclysmic variables or subdwarf B stars.
Interestingly, the prototype eclipsing binary \ebgiant\ does not 
show heartbeat events, although its orbital properties are compatible with the 
sample by \citet{beck13}.

\subsection{Giants in Hierarchical Triple Systems: The Case of HD181068}

An exciting discovery in the early phases of the \kep\ mission was the existence of 
hierarchical triply eclipsing triple systems. The first example, 
presented by \citet{carter12}, consists of three low-mass main-sequence stars and allowed 
a full dynamical solution of all components by measuring eclipse-timing variations, 
without the need for radial-velocity follow-up observations
\citep[a technique that was also applied to confirm numerous multi-planet systems, e.g.][]{fabrycky12}.

\begin{figure}
\begin{center}
\resizebox{\hsize}{!}{\includegraphics{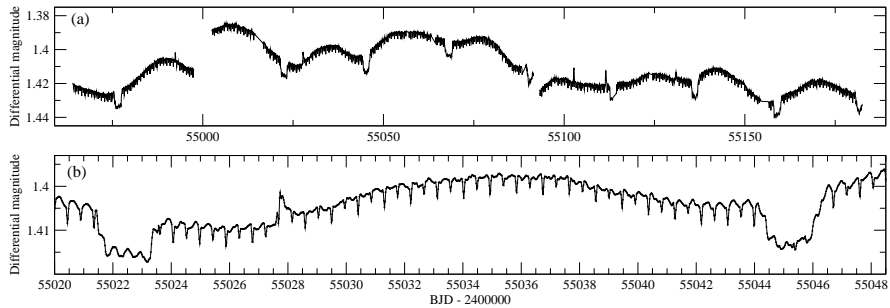}}
\caption{Discovery light curve of HD\,181068, a triply eclipsing triple system consisting of 
a red giant primary and two main-sequence stars. The bottom panel shows a close-up of 
one orbital period with a primary and secondary eclipse, interleaved by $\sim0.9$\,d 
eclipses of the main-sequence binary. From \citet{derekas11}.}
\label{fig:trinity}
\end{center}
\end{figure}

Shortly after, \citet{derekas11} presented the discovery of the first 
triply eclipsing triple system with a red-giant component. 
Figure \ref{fig:trinity} shows the 
discovery light curve of HD\,181068 (also known as ``Trinity''), 
which contains long-duration eclipses with an interval of $\sim$\,23 days 
interleaved by short-duration eclipses with an interval of $\sim$\,0.4 days. 
Follow-up radial velocity and interferometric observations
confirmed that the primary is a red giant which is eclipsed by a pair of main-sequence 
stars with an orbital period of 45.5 days, while the low-mass binary itself 
eclipses every $\sim$\,0.9 days. The short-period 
eclipses disappear during primary and secondary eclipse 
(see Figure \ref{fig:trinity})  
because of the similar temperatures (and hence surface brightnesses) of the 
three components. Subsequent modeling of eclipse timing variations of the 
outer binary yielded a full dynamical solution, with radii and masses of all three 
components measured to better than 5\% \citep{borkovits12}.

\begin{figure}
\begin{center}
\resizebox{\hsize}{!}{\includegraphics{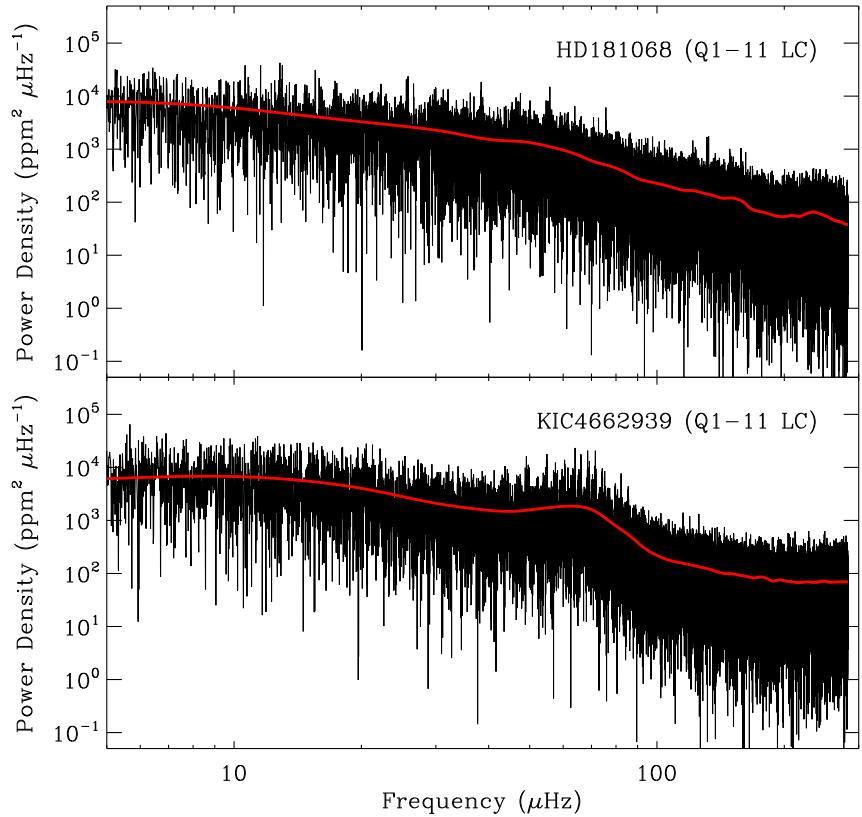}}
\caption{Top panel: Power spectrum of HD\,181068 after removal of the eclipses from the 
\textit{Kepler} light curve. The thick red line shows a heavily smoothed version of the 
spectrum. Bottom panel: Power spectrum of KIC\,4662939, a field red giant with similar 
fundamental properties as HD\,181068\,A. Note the absence of stochastic 
oscillations near 70\,\muHz\ in the top panel. From \citet{fuller13}.}
\label{fig:trinityspec}
\end{center}
\end{figure}

A remarkable aspect of the HD\,181068 system is the lack of stochastic 
oscillations. Figure \ref{fig:trinityspec} compares a power spectrum of 
HD181068 (after removing 
all eclipses) to an oscillating field giant with 
similar fundamental properties as HD181068\,A. Interestingly the 
granulation background, which manifests itself as red noise in the power spectrum, 
is very similar in both stars, while the power excess due to 
oscillations is completely suppressed in HD181068\,A. As speculated by 
\citet{fuller13}, the close dwarf components may be responsible for 
this suppression by tidally synchronizing the rotational frequency of the red giant 
with the long-period orbit, and causing increased magnetic activity which has been 
suggested to suppress the excitation of stochastic oscillations \citep{chaplin11c}. 
HD181068\,A is the first confirmed case of suppressed stochastic oscillations in a binary 
system, and further evidence that this suppression mechanism is indeed related to 
the binary interactions has been recently found for other candidate eclipsing binaries with 
red-giant components \citep{gaulme14}.

While no stochastic oscillations are observed, HD\,181068\,A shows high amplitude 
pulsations at lower frequencies, which can be clearly identified during primary and 
secondary eclipse (Figure \ref{fig:trinity}). The peaks 
in the amplitude spectrum are narrow, 
indicating that they are not stochastically driven, and are linear combinations of the 
long ($\omega_{1}$) and short orbital frequency ($\omega_{23}$) with frequencies of 
$f_{1}=2(\omega_{23}-2\omega_{1})$, $f_{2}=2(\omega_{23}-\omega_{1})$, 
$f_{3}=\omega_{23}$ and $f_{4}=2\omega_{23}$. 
\citet{fuller13} 
demonstrated that these frequencies can be explained by three body tidal forces, 
which cause the orbital motion of the outer pair to induce pulsations in red-giant 
primary.

\section{Dwarf and Subgiant Stars}

Asteroseismology of dwarfs and subgiants in eclipsing binaries has traditionally 
focused on coherent pulsators such as \dscut\ and \gdor\ stars, 
since these show larger amplitudes than stochastic oscillators 
and hence eclipses and pulsations can be more easily detected 
using ground-based observations. At the time of writing of this review 
no stochastic oscillations in a dwarf or subgiant component in an eclipsing binary have 
been published, although at least two \kep\ detections are in preparation 
(Basu et al. \& Sharp et al., in preparation).

\subsection{Classical Pulsators}

Classical pulsations in A and F stars have been successfully detected
using ground-based observations in a few dozen detached and semi-detached 
eclipsing binary systems
\citep[see, e.g.,][and references therein]{mkrtichian04,rodriguez01}. 
However, the short time base and precision of ground-based photometry often 
limited the number of reliable pulsation frequencies that were detected, and 
only for a small number of these systems full dynamical orbits could be combined with a
secure detection of high-amplitude pulsations \citep[see, e.g.,][]{christiansen07}. 

\begin{figure}
\begin{center}
\resizebox{\hsize}{!}{\includegraphics{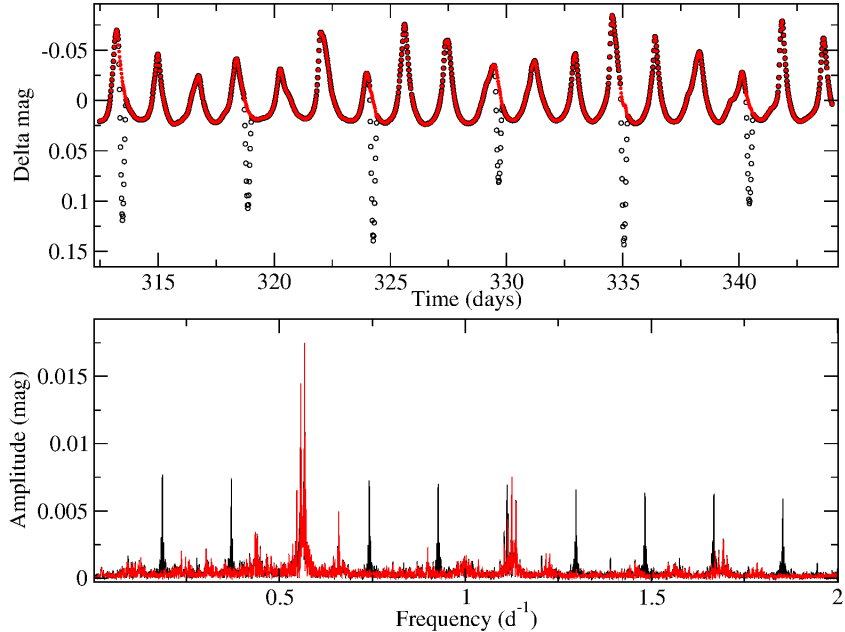}}
\caption{Top panel: Subset of the \textit{Kepler} light curve for KIC\,11285625, a system 
showing eclipses with a period of 10.8 days and \gdor\ pulsations with 
periods of $\sim1-2$\,days. Filled red data points show the 
light curve after removal of the eclipses.
Bottom panel: Amplitude spectrum of the \textit{Kepler} data before (black) and after (red) 
removing the eclipses. From \citet{debosscher13}.}
\label{fig:classeb}
\end{center}
\end{figure}

Similar to stochastic oscillations, CoRoT and \kep\ dramatically changed 
this picture. Following first detections of \gdor\ pulsations in eclipsing binaries 
discovered by CoRoT \citep{damiani10,sokolovsky10}, \citet{maceroni09} presented a 
radial velocity orbit of an eclipsing binary consisting of two B stars, 
with additional variations that could either be attributed 
to self-driven or tidally induced pulsations. Firm detections of self-excited 
\gdor\ pulsations \citep{debosscher13,maceroni13}, \dscut\ pulsations 
\citep{southworth11b,lehmann13}, and hybrid \gdor-\dscut\ pulsations 
\citep{hambleton13,maceroni14} in double-lined spectroscopic and 
eclipsing binaries followed. The studies
illustrated that disentangling the pulsational variability from 
variability induced by the binary orbit requires careful iterative techniques 
(see Figure \ref{fig:classeb}), and 
showed up limitations of current light curve modeling codes to account for 
complex reflection effects \citep[e.g.,][]{southworth11b}.
Each of these studies constrained the primary and secondary masses, radii and 
temperatures with uncertainties of a few percent, and yielded a wealth of pulsation 
frequencies. 

Detailed asteroseismic modeling of these systems is still in progress, and first 
efforts have concentrated on identifying the pulsating components 
by modeling the expected pulsation frequency 
ranges of p modes and g modes based on the dynamically constrained properties. 
For example, \citet{maceroni14} showed that the estimated mean g-mode period spacing 
measured in the F-star eclipsing binary KIC\,3858884 is only consistent with pulsations in the 
secondary, despite both components having similar masses and differing in radius by 
$\sim$10\,\%. Furthermore, inclusion of convective core overshooting was required to 
obtain agreement with theoretical models.
The exceptional amount of 
complementary information in KIC\,3858884 and other systems 
promises to advance our understanding of pulsations in intermediate-mass stars in 
future modeling efforts.

Turning to more evolved stars, ground-based observations have
yielded a handful of RR\,Lyrae and Cepheid pulsators in eclipsing binary systems 
\citep{pietrzynski10,soszynski11}. Such systems are important to test 
masses derived from evolutionary models, 
as well as to accurately measure distances through 
period-luminosity relations. So far, no detections of RR\,Lyrae or Cepheid 
pulsators in eclipsing binaries have been reported using space-based observations, 
which is likely related to the 
relative sparsity of such stars in the \kep\ and CoRoT target lists.

\subsection{Compact Pulsators}

Asteroseismology of white dwarfs or subdwarf B stars 
allows to address a wide variety of fundamental physics such as convection, 
crystallization, the properties of neutrinos, and the evolution
on the extreme horizontal 
branch \citep[see, e.g.,][for reviews]{winget08,heber09}. Finding compact pulsators 
in eclipsing binary systems is extremely valuable to 
cross-check asteroseismically derived 
properties and provide independent contraints for improved seismic modeling. 

Due to their faintness, the detection of pulsating white dwarfs or sdB stars 
in eclipsing binaries is challenging.  
The benchmark system is PG1336-018,
a 0.1 day period eclipsing sdB - M dwarf system \citep{kilkenny98}.
The sdB component 
shows p-mode pulsations with frequencies ranging from $\sim5000-7000$\muHz, which were 
subsequently 
combined with a full orbital solution to demonstrate that asteroseismic modeling yields a mass 
and radius which agrees with the 
dynamical estimates within 1$\%$ \citep{vukovic07,vangrootel13}.

\begin{figure}[t!]
\begin{center}
\resizebox{9cm}{!}{\includegraphics{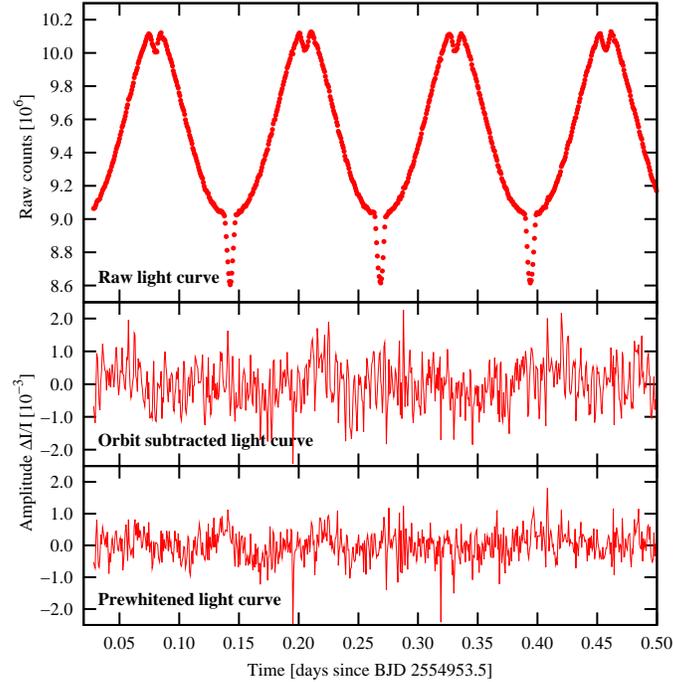}}
\caption{Top panel: \textit{Kepler} light curve of the eclipsing sdB+M binary KIC\,9472174. 
Middle panel: Light curve after subtracting the orbital solution. The 
pulsations in the sdB star are clearly visible. 
Bottom panel: Residuals after subtracting the modeled pulsation frequencies. 
From \citet{ostensen10}.}
\label{fig:sbd}
\end{center}
\end{figure}

A spectacular second detection of a pulsating sdB star in an eclipsing binary has 
been revealed by \kep\ \citep{ostensen10}. KIC\,9472174 (2M1938+4603, $V\sim12.3$) 
has an orbital period of 
0.12 days, with strong variations due to the reflection effect in the light curve 
(Fig \ref{fig:sbd}). 
After removal of a light curve model, the residuals show an unusually rich frequency 
spectrum in the 
sdB component. \citet{ostensen10} report a total of 55 pulsation frequencies spanning 
from $50-4500\muHz$, 
which are attributed to both p-mode and g-mode pulsations. The orbital 
solution combined with the radial velocity semi-amplitude and spectroscopic gravity 
yielded a mass of 
$M=0.48\pm0.03\msun$, consistent with a post common-envelope sdB star. 
Modeling of the pulsation frequencies is 
expected to yield constraints on the core structure and hence the progenitor 
mass of the sdB star.

Apart from KIC\,9472174, \kep\ has also uncovered several pulsating sdB stars in binaries 
showing light variations due to the reflection effect \citep{kawaler10b}, which have been 
used to investigate tidal synchronization timescales by comparing rotation periods 
measured from rotational splittings to the binary period inferred from the light 
curve \citep{pablo12}. Additionally, \kep\ has also uncovered an eclipsing (but 
non-pulsating) sdB + white dwarf binary which shows a combination of binary effects such 
as ellipsoidal deformation, Doppler beaming and microlensing \citep{bloemen11}.

\section{Summary and Future Prospects}

Table \ref{tab:summary} summarizes the characteristics of confirmed eclipsing and/or heartbeat 
systems for which asteroseismic detections (either self-driven or tidally-induced) 
have been made using space-based observations. 
The list illustrates that 
the synergy of asteroseismology and eclipsing/eccentric binary stars using space-based 
observations is (unsurprisingly) 
still in its infancy: all of the discussed systems have been 
published within the last four years. Most detections were made by \kep, 
which provided the required continuous 
monitoring to detect eclipses, and high-precision photometry to detect 
oscillations despite the dilution by the binary component or small amplitudes in the 
pulsating star. 
As noted in Section 4.1, the list of confirmed red giants in eclipsing systems can be 
expected to increase significantly once sufficient radial velocities have been gathered 
for the \kep\ candidate systems \citep{gaulme13}.

\begin{table*}
\begin{small}
\begin{center}
\caption{List of eclipsing and/or heartbeat systems with components showing 
self-excited or tidally-induced pulsations detected from space-based observations. 
Systems are grouped into giants (top), intermediate and high-mass stars (middle) and 
compact stars (bottom). 
Columns list the approximate $V$-band magnitude, spectral types (Sp.Type), 
orbital period ($P$), eccentricity ($e$), and flags indicating 
a double-lined spectroscopic orbit (SB2), stochastic oscillations (stoch.), 
self-excited coherent pulsations, tidal pulsations, eclipses (Ecl.), and heartbeat 
effects (HB).}
\begin{tabular}{l c c c c c c c c c c c}        
\hline 
ID & $V$ & Sp.Type & $P$(d) & $e$ & SB2 & stoch. & coherent & tidal & Ecl. & HB & Ref \\
\hline		
HD\,181068		& 8.0	& KIII+MV+MV& 45.5+0.91	& 0.0	& no	& no	& no	& yes	& yes	& no	& a \\
KIC\,5006817	& 10.9	& KIII+MV	& 94.8		& 0.71	& no	& yes	& no	& no	& no	& yes	& b \\
KIC\,8410637	& 11.3	& KIII+FV	& 408.3		& 0.69	& yes	& yes	& no	& no	& yes	& no	& c \\
\hline
KIC\,10661783	& 9.5	& GIV+AV	& 1.2	& 0.0	& yes	& no	& yes	& no	& yes	& no	& d \\
KIC\,4544587	& 10.8	& FV+FV		& 2.2	& 0.29	& yes	& no	& yes	& yes	& yes	& yes	& e \\
HD\,174884		& 8.4	& BV+BV		& 3.7	& 0.29	& yes	& no	& maybe	& maybe	& yes	& yes	& f \\
CID\,102918586	& 11.7	& FV+FV		& 4.4	& 0.25	& yes	& no	& yes	& yes	& yes	& yes	& g \\
KIC\,11285625	& 10.1	& FV+FV		& 10.8	& 0.0	& yes	& no	& yes	& no	& yes	& no	& h \\
KIC\,3858884	& 9.3	& FV+FV		& 26.0	& 0.47	& yes	& no	& yes	& maybe	& yes	& no	& i	\\	
HD\,187091		& 8.4	& AV+AV		& 41.8	& 0.83	& yes	& no	& no	& yes	& no	& yes	& j \\
\hline
KIC\,9472174	& 12.3	& sdB+MV	& 0.13	& --	& no	& no	& yes	& no	& yes	& no	& k \\
\hline
\end{tabular} 
\label{tab:summary} 
\end{center}
\flushleft References: 
(a) \citet{derekas11,borkovits12,fuller13},
(b) \citet{beck13}, 
(c) \citet{hekker10,frandsen13}, 
(d) \citet{southworth11,lehmann13}, 
(e) \citet{hambleton13}, 
(f) \citet{maceroni09},
(g) \citet{maceroni13} , 
(h) \citet{debosscher13}, 
(i) \citet{maceroni14}, 
(j) \citet{welsh11}, 
(k) \citet{ostensen10}.
\end{small}
\end{table*}

\begin{figure}
\begin{center}
\resizebox{8.3cm}{!}{\includegraphics{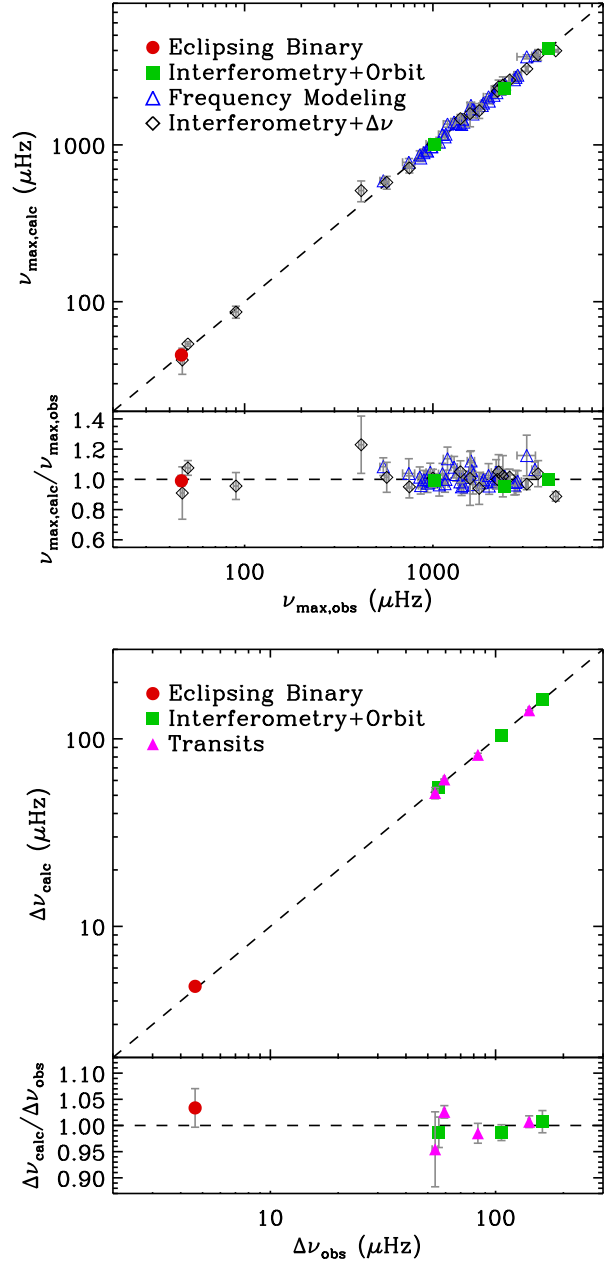}}
\caption{Empirical tests of asteroseismic scaling relations. 
Top: Comparison of measured \numax\ values to empirical values calculated from 
independent measurements (see legend). Filled symbols are  
empirical values which are independent of asteroseismology. Bottom: Same as top 
figure but for the \Dnu\ scaling relation.}
\label{fig:scalingtest}
\end{center}
\end{figure}

Continued observations and the search for new 
asteroseismic eclipsing binaries remains crucial to answer important questions 
regarding stellar pulsations and evolution across the HRD. To highlight one 
particular aspect, Figure \ref{fig:scalingtest} shows updated empirical 
tests of the \numax\ and \Dnu\ scaling relations for stochastic oscillators.
Note that most points in the \numax\ comparison 
are not fully independent from asteroseismology,
with properties determined either 
by combining interferometric angular diameters with asteroseismic densities calculated from 
the \Dnu\ scaling relation \citep[e.g.,][]{huber12b}, or masses and radii determined from  
individual frequency modeling \citep[e.g.,][]{metcalfe14}.
\ebgiant\ marks the first datapoint from an 
eclipsing binary, allowing an
independent test of \numax\ and \Dnu\ that is otherwise only possible for wide 
binaries with radii and masses measured from astrometry and interferometry. 
The \Dnu\ comparison includes four 
stars which host exoplanets with independently constrained 
eccentricities, which allows an independent measurement of the mean stellar density 
\citep{seager03,winn10b}: HD17156 \citep{nutzman11,gilliland11}, 
TrES-2 \citep{barclay12,southworth11}, Hat-P7 \citep{cd10,southworth11}, 
and Kepler-14 \citep{huber13,southworth12}. 
Note that only stars with calculated 
\numax\ and \Dnu\ with uncertainties better than 20\% and 10\% have been included 
in the comparison.

The median residuals for both quantities are close to zero, with a 
scatter of $7$\% for \numax\ and $3$\% for \Dnu. While these numbers are encouraging, 
it is important to note that the observational uncertainties for \numax\ and \Dnu\ 
derived from \kep\ data are typically up to a factor of 2 or more smaller 
\citep{chaplin13}. Additionally, comparisons for evolved giant stars are essentially limited 
to one datapoint, yet the vast majority of stars with asteroseismic detections are giants 
(see Figure \ref{fig:history}). Indeed, \ebgiant\
indicates that 
the \numax\ scaling relation remains accurate for giant stars, while the \Dnu\ scaling 
relation predicts a density which is too low by $\sim$7\%. 
Observations of stochastic oscillations in additional eclipsing binary systems 
will be required to investigate whether this offset may be systematic, and allow 
an \textit{empirical} calibration of scaling relations. 

Future observations of asteroseismic eclipsing binaries with giant and dwarf 
components 
can be expected from space-based missions such as K2 \citep{howell14}, 
TESS \citep{ricker09} and PLATO \citep{rauer13}. 
Importantly, these mission will observe stars 
which are significantly brighter than typical CoRoT and \kep\ targets, 
hence increasing the potential for independent constraints from ground-based 
observations such as long-baseline interferometry. There is little doubt that future  
space-based observations of asteroseismic eclipsing binary stars, combined with 
improved modeling efforts, will continue to play an important role to advance our 
understanding of stellar evolution across the H-R diagram.

\vspace{0.5cm}

\noindent
\textbf{Acknowledgements:} I thank the organizers Elizabeth Griffin and Bob Stencel for a 
fantastic conference, 
and I am grateful to Paul Beck, J{\o}rgen Christensen-Dalsgaard, Orlagh Creevey, 
Jonas Debosscher, Aliz Derekas, Saskia Hekker, S{\o}ren Frandsen, Jim Fuller and 
Roy {\O}stensen for providing figures and comments on the manuscript. 
Financial support was provided by an appointment to the NASA Postdoctoral Program 
at Ames Research Center administered by Oak Ridge Associated Universities, and NASA 
Grant NNX14AB92G issued through the Kepler Participating Scientist Program.

\newpage
\bibliographystyle{assl}
\bibliography{refs}

\begin{thebibliography}{124}
\providecommand{\natexlab}[1]{#1}

\bibitem[{{Aerts}(2013)}]{aerts13}
{Aerts}, C., 2013, in EAS Publications Series, volume~64 of EAS Publications
  Series, pp. 323--330

\bibitem[{{Aerts} et~al.(2010){Aerts}, {Christensen-Dalsgaard}, and
  {Kurtz}}]{aerts10}
{Aerts}, C., {Christensen-Dalsgaard}, J., and {Kurtz}, D.W., 2010,
  {Asteroseismology}, Springer: Dodrecht

\bibitem[{{Antoci} et~al.(2011){Antoci}, {Handler}, {Campante}
  et~al.}]{antoci11}
{Antoci}, V., {Handler}, G., {Campante}, T.L., et~al., 2011, \nat, 477, 570

\bibitem[{{Barclay} et~al.(2012){Barclay}, {Huber}, {Rowe} et~al.}]{barclay12}
{Barclay}, T., {Huber}, D., {Rowe}, J.F., et~al., 2012, \apj, 761, 53

\bibitem[{{Beck} et~al.(2011){Beck}, {Bedding}, {Mosser} et~al.}]{beck11}
{Beck}, P.G., {Bedding}, T.R., {Mosser}, B., et~al., 2011, Science, 332, 205

\bibitem[{{Beck} et~al.(2013){Beck}, {Hambleton}, {Vos} et~al.}]{beck13}
{Beck}, P.G., {Hambleton}, K., {Vos}, J., et~al., 2013, \aap, in press
  (arXiv:1312.4500)

\bibitem[{{Bedding}(2011)}]{bedding11b}
{Bedding}, T.R., 2011, ArXiv e-prints \rm (arXiv:1107.1723)

\bibitem[{{Bedding} et~al.(2010){Bedding}, {Kjeldsen}, {Campante}
  et~al.}]{bedding10}
{Bedding}, T.R., {Kjeldsen}, H., {Campante}, T.L., et~al., 2010, \apj, 713, 935

\bibitem[{{Bedding} et~al.(2011){Bedding}, {Mosser}, {Huber}
  et~al.}]{bedding11}
{Bedding}, T.R., {Mosser}, B., {Huber}, D., et~al., 2011, \nat, 471, 608

\bibitem[{{Belkacem}(2012)}]{belkacem12}
{Belkacem}, K., 2012, in S.~{Boissier}, P.~{de Laverny}, N.~{Nardetto},
  R.~{Samadi}, D.~{Valls-Gabaud}, and H.~{Wozniak}, editors, SF2A-2012:
  Proceedings of the Annual meeting of the French Society of Astronomy and
  Astrophysics, pp. 173--188

\bibitem[{{Belkacem} et~al.(2011){Belkacem}, {Goupil}, {Dupret}
  et~al.}]{belkacem11}
{Belkacem}, K., {Goupil}, M.J., {Dupret}, M.A., et~al., 2011, \aap, 530, A142

\bibitem[{{Bloemen} et~al.(2011){Bloemen}, {Marsh}, {{\O}stensen}
  et~al.}]{bloemen11}
{Bloemen}, S., {Marsh}, T.R., {{\O}stensen}, R.H., et~al., 2011, \mnras, 410,
  1787

\bibitem[{{Borkovits} et~al.(2013){Borkovits}, {Derekas}, {Kiss}
  et~al.}]{borkovits12}
{Borkovits}, T., {Derekas}, A., {Kiss}, L.L., et~al., 2013, \mnras, 428, 1656

\bibitem[{{Bouchy} and {Carrier}(2001)}]{bouchy01}
{Bouchy}, F. and {Carrier}, F., 2001, \aap, 374, L5

\bibitem[{{Breger}(2000)}]{breger00}
{Breger}, M., 2000, in {M.~Breger \& M.~Montgomery}, editor, Delta Scuti and
  Related Stars, volume 210 of Astronomical Society of the Pacific Conference
  Series, p.~3

\bibitem[{{Breger} et~al.(2011){Breger}, {Balona}, {Lenz} et~al.}]{breger11}
{Breger}, M., {Balona}, L., {Lenz}, P., et~al., 2011, \mnras, 414, 1721

\bibitem[{{Brogaard} et~al.(2012){Brogaard}, {VandenBerg}, {Bruntt}
  et~al.}]{brogaard12}
{Brogaard}, K., {VandenBerg}, D.A., {Bruntt}, H., et~al., 2012, \aap, 543, A106

\bibitem[{{Brown} et~al.(1991){Brown}, {Gilliland}, {Noyes} et~al.}]{brown91}
{Brown}, T.M., {Gilliland}, R.L., {Noyes}, R.W., et~al., 1991, \apj, 368, 599

\bibitem[{{Brown} et~al.(2011){Brown}, {Latham}, {Everett} et~al.}]{brown11}
{Brown}, T.M., {Latham}, D.W., {Everett}, M.E., et~al., 2011, \aj, 142, 112

\bibitem[{{Burkart} et~al.(2012){Burkart}, {Quataert}, {Arras}
  et~al.}]{burkart12}
{Burkart}, J., {Quataert}, E., {Arras}, P., et~al., 2012, \mnras, 421, 983

\bibitem[{{Carrier} et~al.(2001){Carrier}, {Bouchy}, {Kienzle}
  et~al.}]{carrier01}
{Carrier}, F., {Bouchy}, F., {Kienzle}, F., et~al., 2001, \aap, 378, 142

\bibitem[{{Carter} et~al.(2012){Carter}, {Agol}, {Chaplin} et~al.}]{carter12}
{Carter}, J.A., {Agol}, E., {Chaplin}, W.J., et~al., 2012, Science, 337, 556

\bibitem[{{Casagrande} et~al.(2014){Casagrande}, {Silva Aguirre}, {Stello}
  et~al.}]{casagrande14}
{Casagrande}, L., {Silva Aguirre}, V., {Stello}, D., et~al., 2014, ArXiv
  e-prints (arXiv:1403.2754)

\bibitem[{{Chaplin} et~al.(2014){Chaplin}, {Basu}, {Huber} et~al.}]{chaplin13}
{Chaplin}, W.J., {Basu}, S., {Huber}, D., et~al., 2014, \apjs, 210, 1

\bibitem[{{Chaplin} et~al.(2011{\natexlab{a}}){Chaplin}, {Bedding}, {Bonanno}
  et~al.}]{chaplin11c}
{Chaplin}, W.J., {Bedding}, T.R., {Bonanno}, A., et~al., 2011{\natexlab{a}},
  \apjl, 732, L5

\bibitem[{{Chaplin} et~al.(2011{\natexlab{b}}){Chaplin}, {Kjeldsen},
  {Christensen-Dalsgaard} et~al.}]{chaplin11a}
{Chaplin}, W.J., {Kjeldsen}, H., {Christensen-Dalsgaard}, J., et~al.,
  2011{\natexlab{b}}, Science, 332, 213

\bibitem[{{Chaplin} and {Miglio}(2013)}]{chaplin13b}
{Chaplin}, W.J. and {Miglio}, A., 2013, \araa, 51, 353

\bibitem[{{Christensen-Dalsgaard}(2003)}]{CD03}
{Christensen-Dalsgaard}, 2003, Lecture Notes, Aarhus University

\bibitem[{{Christensen-Dalsgaard} et~al.(2010){Christensen-Dalsgaard},
  {Kjeldsen}, {Brown} et~al.}]{cd10}
{Christensen-Dalsgaard}, J., {Kjeldsen}, H., {Brown}, T.M., et~al., 2010,
  \apjl, 713, L164

\bibitem[{{Christiansen} et~al.(2007){Christiansen}, {Derekas}, {Ashley}
  et~al.}]{christiansen07}
{Christiansen}, J.L., {Derekas}, A., {Ashley}, M.C.B., et~al., 2007, \mnras,
  382, 239

\bibitem[{{Creevey} et~al.(2011){Creevey}, {Metcalfe}, {Brown}
  et~al.}]{creevey11}
{Creevey}, O.L., {Metcalfe}, T.S., {Brown}, T.M., et~al., 2011, \apj, 733, 38

\bibitem[{{Damiani} et~al.(2010){Damiani}, {Maceroni}, {Cardini}
  et~al.}]{damiani10}
{Damiani}, C., {Maceroni}, C., {Cardini}, D., et~al., 2010, \apss, 328, 91

\bibitem[{{De Ridder} et~al.(2009){De Ridder}, {Barban}, {Baudin}
  et~al.}]{deridder09}
{De Ridder}, J., {Barban}, C., {Baudin}, F., et~al., 2009, \nat, 459, 398

\bibitem[{{Debosscher} et~al.(2013){Debosscher}, {Aerts}, {Tkachenko}
  et~al.}]{debosscher13}
{Debosscher}, J., {Aerts}, C., {Tkachenko}, A., et~al., 2013, \aap, 556, A56

\bibitem[{{Derekas} et~al.(2011){Derekas}, {Kiss}, {Borkovits}
  et~al.}]{derekas11}
{Derekas}, A., {Kiss}, L.L., {Borkovits}, T., et~al., 2011, Science, 332, 216

\bibitem[{Dziembowski et~al.(2001)Dziembowski, {Gough}, {Houdek}
  et~al.}]{dziembowski01}
Dziembowski, W.A., {Gough}, D.O., {Houdek}, G., et~al., 2001, \mnras, 328, 601

\bibitem[{{Fabrycky} et~al.(2012){Fabrycky}, {Ford}, {Steffen}
  et~al.}]{fabrycky12}
{Fabrycky}, D.C., {Ford}, E.B., {Steffen}, J.H., et~al., 2012, \apj, 750, 114

\bibitem[{{Frandsen} et~al.(2013){Frandsen}, {Lehmann}, {Hekker}
  et~al.}]{frandsen13}
{Frandsen}, S., {Lehmann}, H., {Hekker}, S., et~al., 2013, \aap, 556, A138

\bibitem[{{Fuller} et~al.(2013){Fuller}, {Derekas}, {Borkovits}
  et~al.}]{fuller13}
{Fuller}, J., {Derekas}, A., {Borkovits}, T., et~al., 2013, \mnras, 429, 2425

\bibitem[{{Fuller} and {Lai}(2012)}]{fuller12}
{Fuller}, J. and {Lai}, D., 2012, \mnras, 420, 3126

\bibitem[{{Gaulme} et~al.(2014){Gaulme}, {Jackiewicz}, {Appourchaux}
  et~al.}]{gaulme14}
{Gaulme}, P., {Jackiewicz}, J., {Appourchaux}, T., et~al., 2014, \apj, 785, 5

\bibitem[{{Gaulme} et~al.(2013){Gaulme}, {McKeever}, {Rawls} et~al.}]{gaulme13}
{Gaulme}, P., {McKeever}, J., {Rawls}, M.L., et~al., 2013, \apj, 767, 82

\bibitem[{{Gilliland} et~al.(2010){Gilliland}, {Brown}, {Christensen-Dalsgaard}
  et~al.}]{gilliland10}
{Gilliland}, R.L., {Brown}, T.M., {Christensen-Dalsgaard}, J., et~al., 2010,
  \pasp, 122, 131

\bibitem[{{Gilliland} et~al.(2011){Gilliland}, {McCullough}, {Nelan}
  et~al.}]{gilliland11}
{Gilliland}, R.L., {McCullough}, P.R., {Nelan}, E.P., et~al., 2011, \apj, 726,
  2

\bibitem[{{Gizon} and {Solanki}(2003)}]{gizon03}
{Gizon}, L. and {Solanki}, S.K., 2003, \apj, 589, 1009

\bibitem[{Gough(1986)}]{gough86}
Gough, D.O., 1986, in Y.~{Osaki}, editor, Hydrodynamic and Magnetodynamic
  Problems in the Sun and Stars, p. 117, Uni. of Tokyo Press

\bibitem[{{Grigahc{\`e}ne} et~al.(2010){Grigahc{\`e}ne}, {Antoci}, {Balona}
  et~al.}]{grigahcene10}
{Grigahc{\`e}ne}, A., {Antoci}, V., {Balona}, L., et~al., 2010, \apjl, 713,
  L192

\bibitem[{{Guenther} et~al.(2007){Guenther}, {Kallinger}, {Reegen}
  et~al.}]{guenther07}
{Guenther}, D.B., {Kallinger}, T., {Reegen}, P., et~al., 2007, Communications
  in Asteroseismology, 151, 5

\bibitem[{{Guzik} et~al.(2000){Guzik}, {Kaye}, {Bradley} et~al.}]{guzik00}
{Guzik}, J.A., {Kaye}, A.B., {Bradley}, P.A., et~al., 2000, \apjl, 542, L57

\bibitem[{{Hambleton} et~al.(2013){Hambleton}, {Kurtz}, {Pr{\v s}a}
  et~al.}]{hambleton13}
{Hambleton}, K.M., {Kurtz}, D.W., {Pr{\v s}a}, A., et~al., 2013, \mnras, 434,
  925

\bibitem[{{Handler}(2013)}]{handler13}
{Handler}, G., 2013, {Asteroseismology}, p. 207

\bibitem[{{Heber}(2009)}]{heber09}
{Heber}, U., 2009, \araa, 47, 211

\bibitem[{{Hekker} et~al.(2010){Hekker}, {Debosscher}, {Huber}
  et~al.}]{hekker10}
{Hekker}, S., {Debosscher}, J., {Huber}, D., et~al., 2010, \apjl, 713, L187

\bibitem[{{Hekker} et~al.(2013){Hekker}, {Elsworth}, {Basu} et~al.}]{hekker13b}
{Hekker}, S., {Elsworth}, Y., {Basu}, S., et~al., 2013, \mnras, 434, 1668

\bibitem[{{Hekker} et~al.(2012){Hekker}, {Elsworth}, {Mosser}
  et~al.}]{hekker12}
{Hekker}, S., {Elsworth}, Y., {Mosser}, B., et~al., 2012, \aap, 544, A90

\bibitem[{{Hekker} et~al.(2009){Hekker}, {Kallinger}, {Baudin}
  et~al.}]{hekker09}
{Hekker}, S., {Kallinger}, T., {Baudin}, F., et~al., 2009, \aap, 506, 465

\bibitem[{Houdek et~al.(1999)Houdek, {Balmforth}, {Christensen-Dalsgaard}
  et~al.}]{houdek99}
Houdek, G., {Balmforth}, N.J., {Christensen-Dalsgaard}, J., et~al., 1999, \aap,
  351, 582

\bibitem[{{Howell} et~al.(2014){Howell}, {Sobeck}, {Haas} et~al.}]{howell14}
{Howell}, S.B., {Sobeck}, C., {Haas}, M., et~al., 2014, PASP, in press
  (arXiv:1402.5163)

\bibitem[{{Huber} et~al.(2011){Huber}, {Bedding}, {Stello} et~al.}]{huber11b}
{Huber}, D., {Bedding}, T.R., {Stello}, D., et~al., 2011, \apj, 743, 143

\bibitem[{{Huber} et~al.(2013){Huber}, {Chaplin}, {Christensen-Dalsgaard}
  et~al.}]{huber13}
{Huber}, D., {Chaplin}, W.J., {Christensen-Dalsgaard}, J., et~al., 2013, \apj,
  767, 127

\bibitem[{{Huber} et~al.(2012){Huber}, {Ireland}, {Bedding} et~al.}]{huber12b}
{Huber}, D., {Ireland}, M.J., {Bedding}, T.R., et~al., 2012, \apj, 760, 32

\bibitem[{{Huber} et~al.(2014){Huber}, {Silva Aguirre}, {Matthews}
  et~al.}]{huber14}
{Huber}, D., {Silva Aguirre}, V., {Matthews}, J.M., et~al., 2014, \apjs, 211, 2

\bibitem[{{Huber} et~al.(2009){Huber}, {Stello}, {Bedding} et~al.}]{huber09}
{Huber}, D., {Stello}, D., {Bedding}, T.R., et~al., 2009, Communications in
  Asteroseismology, 160, 74

\bibitem[{{Kawaler} et~al.(2010){Kawaler}, {Reed}, {{\O}stensen}
  et~al.}]{kawaler10b}
{Kawaler}, S.D., {Reed}, M.D., {{\O}stensen}, R.H., et~al., 2010, \mnras, 409,
  1509

\bibitem[{{Kilkenny} et~al.(1998){Kilkenny}, {O'Donoghue}, {Koen}
  et~al.}]{kilkenny98}
{Kilkenny}, D., {O'Donoghue}, D., {Koen}, C., et~al., 1998, \mnras, 296, 329

\bibitem[{{Kippenhahn} and {Weigert}(1994)}]{KW}
{Kippenhahn}, R. and {Weigert}, A., 1994, {Stellar Structure and Evolution},
  Springer: Berlin

\bibitem[{{Kjeldsen} and {Bedding}(1995)}]{KB95}
{Kjeldsen}, H. and {Bedding}, T.R., 1995, \aap, 293, 87

\bibitem[{{Kjeldsen} et~al.(2005){Kjeldsen}, {Bedding}, {Butler}
  et~al.}]{kjeldsen05}
{Kjeldsen}, H., {Bedding}, T.R., {Butler}, R.P., et~al., 2005, \apj, 635, 1281

\bibitem[{{Kurtz}(1982)}]{kurtz82}
{Kurtz}, D.W., 1982, \mnras, 200, 807

\bibitem[{{Lai}(1997)}]{lai97}
{Lai}, D., 1997, \apj, 490, 847

\bibitem[{{Lehmann} et~al.(2013){Lehmann}, {Southworth}, {Tkachenko}
  et~al.}]{lehmann13}
{Lehmann}, H., {Southworth}, J., {Tkachenko}, A., et~al., 2013, \aap, 557, A79

\bibitem[{{Lehmann} et~al.(2011){Lehmann}, {Tkachenko}, {Semaan}
  et~al.}]{lehmann11}
{Lehmann}, H., {Tkachenko}, A., {Semaan}, T., et~al., 2011, \aap, 526, A124

\bibitem[{{Maceroni} et~al.(2014){Maceroni}, {Lehmann}, {da Silva}
  et~al.}]{maceroni14}
{Maceroni}, C., {Lehmann}, H., {da Silva}, R., et~al., 2014, ArXiv e-prints

\bibitem[{{Maceroni} et~al.(2013){Maceroni}, {Montalb{\'a}n}, {Gandolfi}
  et~al.}]{maceroni13}
{Maceroni}, C., {Montalb{\'a}n}, J., {Gandolfi}, D., et~al., 2013, \aap, 552,
  A60

\bibitem[{{Maceroni} et~al.(2009){Maceroni}, {Montalb{\'a}n}, {Michel}
  et~al.}]{maceroni09}
{Maceroni}, C., {Montalb{\'a}n}, J., {Michel}, E., et~al., 2009, \aap, 508,
  1375

\bibitem[{{Matijevi{\v c}} et~al.(2012){Matijevi{\v c}}, {Pr{\v s}a}, {Orosz}
  et~al.}]{matijevic12}
{Matijevi{\v c}}, G., {Pr{\v s}a}, A., {Orosz}, J.A., et~al., 2012, \aj, 143,
  123

\bibitem[{{Matthews}(2007)}]{matthews07}
{Matthews}, J.M., 2007, Communications in Asteroseismology, 150, 333

\bibitem[{{M{\'e}sz{\'a}ros} et~al.(2013){M{\'e}sz{\'a}ros}, {Holtzman},
  {Garc{\'{\i}}a P{\'e}rez} et~al.}]{meszaros13}
{M{\'e}sz{\'a}ros}, S., {Holtzman}, J., {Garc{\'{\i}}a P{\'e}rez}, A.E.,
  et~al., 2013, \aj, 146, 133

\bibitem[{{Metcalfe} et~al.(2012){Metcalfe}, {Chaplin}, {Appourchaux}
  et~al.}]{metcalfe12}
{Metcalfe}, T.S., {Chaplin}, W.J., {Appourchaux}, T., et~al., 2012, \apjl, 748,
  L10

\bibitem[{{Metcalfe} et~al.(2014){Metcalfe}, {Creevey}, {Dogan}
  et~al.}]{metcalfe14}
{Metcalfe}, T.S., {Creevey}, O.L., {Dogan}, G., et~al., 2014, ArXiv e-prints
  (arXiv:1402.3614)

\bibitem[{{Michel} and {Baglin}(2012)}]{michel12}
{Michel}, E. and {Baglin}, A., 2012, ArXiv e-prints (arXiv:1202.1422)

\bibitem[{{Michel} et~al.(2009){Michel}, {Samadi}, {Baudin} et~al.}]{michel09}
{Michel}, E., {Samadi}, R., {Baudin}, F., et~al., 2009, \aap, 495, 979

\bibitem[{{Miglio}(2012)}]{miglio11}
{Miglio}, A., 2012, in A.~Miglio, J.~Montalb{\'a}n, and A.~Noels, editors, Red
  Giants as Probes of the Structure and Evolution of the Milky Way, ApSS
  Proceedings, Berlin: Springer

\bibitem[{{Miglio} et~al.(2012){Miglio}, {Brogaard}, {Stello}
  et~al.}]{miglio12b}
{Miglio}, A., {Brogaard}, K., {Stello}, D., et~al., 2012, \mnras, 419, 2077

\bibitem[{{Miglio} et~al.(2013){Miglio}, {Chiappini}, {Morel}
  et~al.}]{miglio13}
{Miglio}, A., {Chiappini}, C., {Morel}, T., et~al., 2013, in European Physical
  Journal Web of Conferences, volume~43 of European Physical Journal Web of
  Conferences, pp. 3004, DOI: 10.1051/epjconf/20134303004

\bibitem[{{Mkrtichian} et~al.(2004){Mkrtichian}, {Kusakin}, {Rodriguez}
  et~al.}]{mkrtichian04}
{Mkrtichian}, D.E., {Kusakin}, A.V., {Rodriguez}, E., et~al., 2004, \aap, 419,
  1015

\bibitem[{{Mosser} et~al.(2010){Mosser}, {Belkacem}, {Goupil}
  et~al.}]{mosser10}
{Mosser}, B., {Belkacem}, K., {Goupil}, M.J., et~al., 2010, \aap, 517, A22

\bibitem[{{Mosser} et~al.(2013){Mosser}, {Michel}, {Belkacem}
  et~al.}]{mosser13}
{Mosser}, B., {Michel}, E., {Belkacem}, K., et~al., 2013, \aap, 550, A126

\bibitem[{{Nemec} et~al.(2013){Nemec}, {Cohen}, {Ripepi} et~al.}]{nemec13}
{Nemec}, J.M., {Cohen}, J.G., {Ripepi}, V., et~al., 2013, \apj, 773, 181

\bibitem[{{Nutzman} et~al.(2011){Nutzman}, {Gilliland}, {McCullough}
  et~al.}]{nutzman11}
{Nutzman}, P., {Gilliland}, R.L., {McCullough}, P.R., et~al., 2011, \apj, 726,
  3

\bibitem[{{{\O}stensen} et~al.(2011){{\O}stensen}, {Bloemen}, {Vu{\v
  c}kovi{\'c}} et~al.}]{ostensen11}
{{\O}stensen}, R.H., {Bloemen}, S., {Vu{\v c}kovi{\'c}}, M., et~al., 2011,
  \apjl, 736, L39

\bibitem[{{{\O}stensen} et~al.(2010){{\O}stensen}, {Green}, {Bloemen}
  et~al.}]{ostensen10}
{{\O}stensen}, R.H., {Green}, E.M., {Bloemen}, S., et~al., 2010, \mnras, 408,
  L51

\bibitem[{{Pablo} et~al.(2012){Pablo}, {Kawaler}, {Reed} et~al.}]{pablo12}
{Pablo}, H., {Kawaler}, S.D., {Reed}, M.D., et~al., 2012, \mnras, 422, 1343

\bibitem[{{Pamyatnykh}(2000)}]{pamyatnykh00}
{Pamyatnykh}, A.A., 2000, in {M.~Breger \& M.~Montgomery}, editor, Delta Scuti
  and Related Stars, volume 210 of Astronomical Society of the Pacific
  Conference Series, p. 215

\bibitem[{{Pietrinferni} et~al.(2004){Pietrinferni}, {Cassisi}, {Salaris}
  et~al.}]{basti}
{Pietrinferni}, A., {Cassisi}, S., {Salaris}, M., et~al., 2004, \apj, 612, 168

\bibitem[{{Pietrzy{\'n}ski} et~al.(2010){Pietrzy{\'n}ski}, {Thompson}, {Gieren}
  et~al.}]{pietrzynski10}
{Pietrzy{\'n}ski}, G., {Thompson}, I.B., {Gieren}, W., et~al., 2010, \nat, 468,
  542

\bibitem[{{Pr{\v s}a} et~al.(2011){Pr{\v s}a}, {Batalha}, {Slawson}
  et~al.}]{prsa11}
{Pr{\v s}a}, A., {Batalha}, N., {Slawson}, R.W., et~al., 2011, \aj, 141, 83

\bibitem[{{Rauer} et~al.(2013){Rauer}, {Catala}, {Aerts} et~al.}]{rauer13}
{Rauer}, H., {Catala}, C., {Aerts}, C., et~al., 2013, ArXiv e-prints
  (arXiv:1310.0696)

\bibitem[{{Ricker} et~al.(2009){Ricker}, {Latham}, {Vanderspek}
  et~al.}]{ricker09}
{Ricker}, G.R., {Latham}, D.W., {Vanderspek}, R.K., et~al., 2009, in American
  Astronomical Society Meeting Abstracts 213, volume~41 of Bulletin of the
  American Astronomical Society, p. 403.01

\bibitem[{{Rodr{\'{\i}}guez} and {Breger}(2001)}]{rodriguez01}
{Rodr{\'{\i}}guez}, E. and {Breger}, M., 2001, \aap, 366, 178

\bibitem[{{Salaris} et~al.(2010){Salaris}, {Cassisi}, {Pietrinferni}
  et~al.}]{salaris10}
{Salaris}, M., {Cassisi}, S., {Pietrinferni}, A., et~al., 2010, \apj, 716, 1241

\bibitem[{{Seager} and {Mall{\'e}n-Ornelas}(2003)}]{seager03}
{Seager}, S. and {Mall{\'e}n-Ornelas}, G., 2003, \apj, 585, 1038

\bibitem[{{Silva Aguirre} et~al.(2012){Silva Aguirre}, {Casagrande}, {Basu}
  et~al.}]{silva12}
{Silva Aguirre}, V., {Casagrande}, L., {Basu}, S., et~al., 2012, \apj, 757, 99

\bibitem[{{Slawson} et~al.(2011){Slawson}, {Pr{\v s}a}, {Welsh}
  et~al.}]{slawson11}
{Slawson}, R.W., {Pr{\v s}a}, A., {Welsh}, W.F., et~al., 2011, \aj, 142, 160

\bibitem[{{Sokolovsky} et~al.(2010){Sokolovsky}, {Maceroni}, {Hareter}
  et~al.}]{sokolovsky10}
{Sokolovsky}, K., {Maceroni}, C., {Hareter}, M., et~al., 2010, Communications
  in Asteroseismology, 161, 55

\bibitem[{{Soszy{\'n}ski} et~al.(2011){Soszy{\'n}ski}, {Dziembowski}, {Udalski}
  et~al.}]{soszynski11}
{Soszy{\'n}ski}, I., {Dziembowski}, W.A., {Udalski}, A., et~al., 2011, \actaa,
  61, 1

\bibitem[{{Southworth}(2011)}]{southworth11}
{Southworth}, J., 2011, \mnras, 417, 2166

\bibitem[{{Southworth}(2012)}]{southworth12}
{Southworth}, J., 2012, \mnras, 426, 1291

\bibitem[{{Southworth} et~al.(2011){Southworth}, {Zima}, {Aerts}
  et~al.}]{southworth11b}
{Southworth}, J., {Zima}, W., {Aerts}, C., et~al., 2011, \mnras, 414, 2413

\bibitem[{{Stello} et~al.(2008){Stello}, {Bruntt}, {Preston} et~al.}]{stello08}
{Stello}, D., {Bruntt}, H., {Preston}, H., et~al., 2008, \apjl, 674, L53

\bibitem[{{Stello} et~al.(2009){Stello}, {Chaplin}, {Bruntt} et~al.}]{stello09}
{Stello}, D., {Chaplin}, W.J., {Bruntt}, H., et~al., 2009, \apj, 700, 1589

\bibitem[{{Stello} et~al.(2013){Stello}, {Huber}, {Bedding} et~al.}]{stello13}
{Stello}, D., {Huber}, D., {Bedding}, T.R., et~al., 2013, \apjl, 765, L41

\bibitem[{Tassoul(1980)}]{tassoul80}
Tassoul, M., 1980, \apjs, 43, 469

\bibitem[{{Thompson} et~al.(2012){Thompson}, {Everett}, {Mullally}
  et~al.}]{thompson12}
{Thompson}, S.E., {Everett}, M., {Mullally}, F., et~al., 2012, \apj, 753, 86

\bibitem[{{Ulrich}(1986)}]{ulrich86}
{Ulrich}, R.K., 1986, \apjl, 306, L37

\bibitem[{{Van Grootel} et~al.(2013){Van Grootel}, {Charpinet}, {Brassard}
  et~al.}]{vangrootel13}
{Van Grootel}, V., {Charpinet}, S., {Brassard}, P., et~al., 2013, \aap, 553,
  A97

\bibitem[{{Vandakurov}(1968)}]{vandakurov68}
{Vandakurov}, Y.V., 1968, \sovast, 11, 630

\bibitem[{{Vu{\v c}kovi{\'c}} et~al.(2007){Vu{\v c}kovi{\'c}}, {Aerts},
  {{\O}stensen} et~al.}]{vukovic07}
{Vu{\v c}kovi{\'c}}, M., {Aerts}, C., {{\O}stensen}, R., et~al., 2007, \aap,
  471, 605

\bibitem[{{Welsh} et~al.(2011){Welsh}, {Orosz}, {Aerts} et~al.}]{welsh11}
{Welsh}, W.F., {Orosz}, J.A., {Aerts}, C., et~al., 2011, \apjs, 197, 4

\bibitem[{{White} et~al.(2011){White}, {Bedding}, {Stello} et~al.}]{white11}
{White}, T.R., {Bedding}, T.R., {Stello}, D., et~al., 2011, \apj, 743, 161

\bibitem[{{Winget} and {Kepler}(2008)}]{winget08}
{Winget}, D.E. and {Kepler}, S.O., 2008, \araa, 46, 157

\bibitem[{{Winn}(2010)}]{winn10b}
{Winn}, J.N., 2010, ArXiv e-prints (arXiv:1001.2010)

\bibitem[{{Winn} et~al.(2010){Winn}, {Fabrycky}, {Albrecht} et~al.}]{winn10}
{Winn}, J.N., {Fabrycky}, D., {Albrecht}, S., et~al., 2010, \apjl, 718, L145

\bibitem[{{Zahn}(1975)}]{zahn75}
{Zahn}, J.P., 1975, \aap, 41, 329

\end{thebibliography}

\end{document}